\documentclass[10pt,twocolumn,letterpaper]{article}

\usepackage{iccv}
\usepackage{times}
\usepackage{epsfig}
\usepackage{graphicx}
\usepackage{amsmath}
\usepackage{amssymb}
\usepackage{paralist}
\usepackage{enumitem}

\usepackage[ruled,vlined]{algorithm2e}
\usepackage{algcompatible}

\usepackage{subfigure}
\usepackage{float}
\usepackage{booktabs}
\usepackage{dblfloatfix}
\usepackage{multirow}

\usepackage{xcolor}
\definecolor{dgreen}{rgb}{0.0, 0.5, 0.0}

\def\xx{\mathtt{x}}
\def\yy{\mathtt{y}}
\def\uu{\mathtt{u}}
\def\vv{\mathtt{v}}

\usepackage[pagebackref=true,breaklinks=true,letterpaper=true,colorlinks,bookmarks=false]{hyperref}

\iccvfinalcopy 

\def\iccvPaperID{2222} 

\ificcvfinal\fi

\begin{document}

\title{Global Adaptive Filtering Layer for Computer Vision}

\author{\\
Viktor Shipitsin\thanks{Contributed equally.},~~~Iaroslav Bespalov\footnotemark[1],~~~Dmitry V. Dylov\thanks{Corresponding author.}\\
Skolkovo Institute of Science and Technology\\
Bolshoy blvd., 30/1, Moscow, Russia 121205\\
{\tt\small \{Viktor.Shipitsin, Iaroslav.Bespalov, d.dylov\}@skoltech.ru}
}
\maketitle
\ificcvfinal\thispagestyle{empty}\fi

\begin{abstract}
We devise a universal adaptive neural layer to ``learn'' optimal frequency filter for each image together with the weights of the base neural network that performs some computer vision task. The proposed approach takes the source image in the spatial domain, automatically selects the best frequencies from the frequency domain, and transmits the inverse-transform image to the main neural network. Remarkably, such a simple add-on layer dramatically improves the performance of the main network regardless of its design. We observe that the light networks gain a noticeable boost in the performance metrics; whereas, the training of the heavy ones converges faster when our adaptive layer is allowed to ``learn'' alongside the main architecture. We validate the idea in four classical computer vision tasks: classification, segmentation, denoising, and erasing, considering popular natural and medical data benchmarks.
%
\end{abstract}

\section{Introduction}
In recent years, computer vision (CV) algorithms have advanced significantly thanks to the advent of artificial neural networks (ANN) and to the development of the computational resources capable of working with them~\cite{szeliski2011computer}. 
At the same time, the constantly growing volumes of data instigated a wave of research efforts involving large neural networks with a colossal number of parameters~\cite{shazeer2017outrageously}, triggering the development of approaches for efficient data processing, model optimization, and training.
One promising trend is not to keep complicating the architectures, but to develop \textit{efficient modules} that allow one to look at computer tasks from a different angle, extracting semantics from the images and ultimately demanding less effort~\cite{GUOEfficient201627,RhuEfficient}. What could be done with the images even \textit{before} they enter a certain neural network is generally concerned with the task of image preprocessing~\cite{Bow-Preprocessing} and will be the leitmotif in this work. 
Particularly, we are interested in developing a `smart' preprocessing module to the following four classical CV problems: segmentation, classification, erasing, and denoising.

The \textit{segmentation} task is one of the most popular tasks in the field of CV, as it allows to localize the object of interest in the image. When image segmentation is concerned, one naturally starts with the U-Net encoder-decoder like models~\cite{Ronneberger}. At the moment, there are various modifications that prove more accurate than the baseline U-Net in various scenarios: Attention U-Net~\cite{Oktay}, U-Net++~\cite{Zhou_1,  Zhou_2}, U-Net 3+~\cite{Huang}, \textit{etc.}. Although much heavier and slower in training, ResNet and DenseNet models are also frequently employed for the purpose of segmentation~\cite{Densenet, Resnet}. One naturally looks for lighter models that would reach the performance level of the heavier models with dozens of millions of parameters~\cite{Resnet,szeliski2011computer}.

Image \textit{classification} is frequently defined as the task of categorizing images into one of several predefined classes, and it is another popular problem in CV~\cite{Waseem-ImageClassification,szeliski2011computer}.
Binary classification is a precursor problem to many other CV challenges, and is an analogy to the segmentation, with the output being a single pixel.
The same segmentation encoders can be employed for the classification problem to obtain an \textit{embedding}, and then, linear layers would predict the class~\cite{Waseem-ImageClassification}.

\textit{Denoising} is another important task in imaging which covers extensive range of domains and applications~\cite{mccann2017convolutional}. Popular approaches, such as DnCNN ~\cite{DnCNN1, DnCNN2}, already became classic and can restore blurred, damaged, and noisy images exceptionally well. 
Naturally, denoising is also the problem where frequency decomposition of an image becomes a particularly important entity for a computer scientist~\cite{Zhang_FSIM,mccann2017convolutional}. It is interesting how the frequency spectra change in the denoising tasks. For example, one can cut out an area from an image and see how a denoising model would paint over the area in the presence of \textit{frequency filtering}. Such erasing~\cite{szeliski2011computer} task will be also briefly considered in this work and is adjacent to another popular problem of \textit{super-resolution} in CV, where the missing pixels are processed to minimize the damage to the image or to maximize a value function such as the resolution.  

The problem of frequency filtering for denoising has been studied very thoroughly in the signal processing and in the imaging physics communities~\cite{Chaudhary,chowdhury2017blood}. In fact, the filtering is at the core of one of the most frequently used clinical imaging modalities -- the ultrasound~\cite{Ihnatsenka}. 
Its typical high\textbackslash mid-range (5-15 MHz) and low (2-5 MHz) frequency probes provide either good resolution or good penetration, but not both at once. The resulting images, therefore, are extremely sensitive to the frequency tuning, with various phenomena such as reverberation, shadowing, excessive absorbance, reflection, and echo, giving the images the distinct grainy look~\cite{Song, Wang}.

What operators of the medical ultrasound do with their knobs on the machine's panel to enhance the appearance of the images in real time has motivated us to mimic the similar `live' filtering for the CV problems. Specifically, we asked ourselves, what if a pre-training block of a neural network would be capable of learning the optimal frequencies for each image in the dataset \textit{live} during the training, effectively maximizing a value function of interest for the entire model? Can we design a universal adaptive layer that would provide the necessary frequency filter for any input image regardless of the network architecture or the CV task at hand? Herein, we present such a solution.

Using the direct and the inverse Fourier transforms, we can switch from the representation of the image in the spatial domain to that in the frequency domain and vice versa.
In the frequency spectrum, particular frequencies are responsible for different properties of the image~\cite{szeliski2011computer}, which can be either enhanced or suppressed with filtering, depending on the value function of interest in one of the four CV problems described above.
For example, the high-pass filter, used for the edge detection, can enhance edges and details, effectively holding promise for improving the segmentation performance if it partook in the training routine along with the main segmentation network. 
In this article, we devise a simple adaptive add-on layer that improves the quality and efficiency of popular neural networks in CV.
The layer learns to automatically find a global filter that will leave only those frequencies that could boost the target metric in the entire dataset (\textit{e.g.}, Dice score in the segmentation, or F$_{1}$-score in the classification problem). 

The rest of the paper is structured as follows. After covering the work related to learning in the Fourier domain in Section \ref{sec:relwork}, we describe the algorithm behind the global adaptive layer in Section~\ref{section:method}. 
We then describe the datasets in Subsection~\ref{section:datasets}: two medical (ultrasound, which has motivated our work) and four natural image data benchmarks, hypothesizing that the wave nature of the ultrasonic data would correspond to a more efficient filtering than that of the natural images.
However, the rest of Section~\ref{section:experiments} reports a likewise enhancement of the baseline performance for the natural images as well. Section~\ref{section:experiments} also reports faster training convergence for the majority of models and tasks and summarizes the results of a controlled and a large-scale studies. 
Sections~\ref{section:discussions} and~\ref{section:conclusions} conclude the paper.
\medskip

\noindent Contributions of this paper are the following:
\begin{compactitem}
\item The first \textit{adaptive layer} to be trained alongside the main neural network to boost its performance by finding \textit{globally optimal} filtering frequencies.
\item A simple, universal, flexible, and intuitive solution for improving and accelerating neural networks.
\item We show at least 6 \% increase in Dice score for light U-Net-like architectures, and accelerate convergence of heavier models (such as ResNet and DenseNet).
\item We report 88 experimental scenarios, 5 variations of the adaptive layer, adding them on to 5 popular architectures, and testing the outcomes on 6 (4 natural and 2 medical) dataset benchmarks in 4 CV tasks.
\item Careful control and large-scale studies are reported.
\end{compactitem}

\section{Related Work}
\label{sec:relwork}
Adoption of the learning algorithms from the \textit{non-image} domains to improve either the target metric or the efficiency of neural networks is a rather recent trait.
The Fourier space is one of such domains, where there are several works reporting the spectral transforms with consequent feature extraction to train their models~\cite{FCNN, MLWCNN, FourierDomainTraining}. Fourier analysis has also been successfully used for dynamic structure segmentation problems, where dynamic structures were distinguished using only the phase spectra~\cite{Li}. We obviously omit a long list of works here, where the frequency data was used for feature engineering or for some domain-specific machine learning applications.

In 2020, however, there appeared a relevant work reporting semantic segmentation with domain adaptation~\cite{Yang}, where the spectral amplitudes of the source and the target images were combined to boost the performance of the model. High-frequency low-dimensional regression problems (where Fourier features improved the results of the coordinate-based multi-layer perceptions for image regression), 3D shape regression, MRI reconstruction, and inverse rendering tasks are also some very recent results~\cite{Tancik}.

These aforementioned works have shown their effectiveness by proving that some information could be lost if one relies merely on the spatial image domain. The difficulty, however, is that the manual selection of the correct frequencies for optimizing an ANN is not a simple task. Remarkably, none of the algorithms makes effort to optimize the frequency spectra during the training routine of the main architecture.
\textit{Therefore, our solution is to automate the search for the optimal weights in the frequency spectrum until the desired metric of a given network is maximized for each CV task.} 
Despite being rather intuitive, such a solution has not been reported in the literature, motivating our study herein.

\section{Method}
\label{section:method}

\begin{figure}[t]
\begin{center}
\includegraphics[width=\columnwidth]{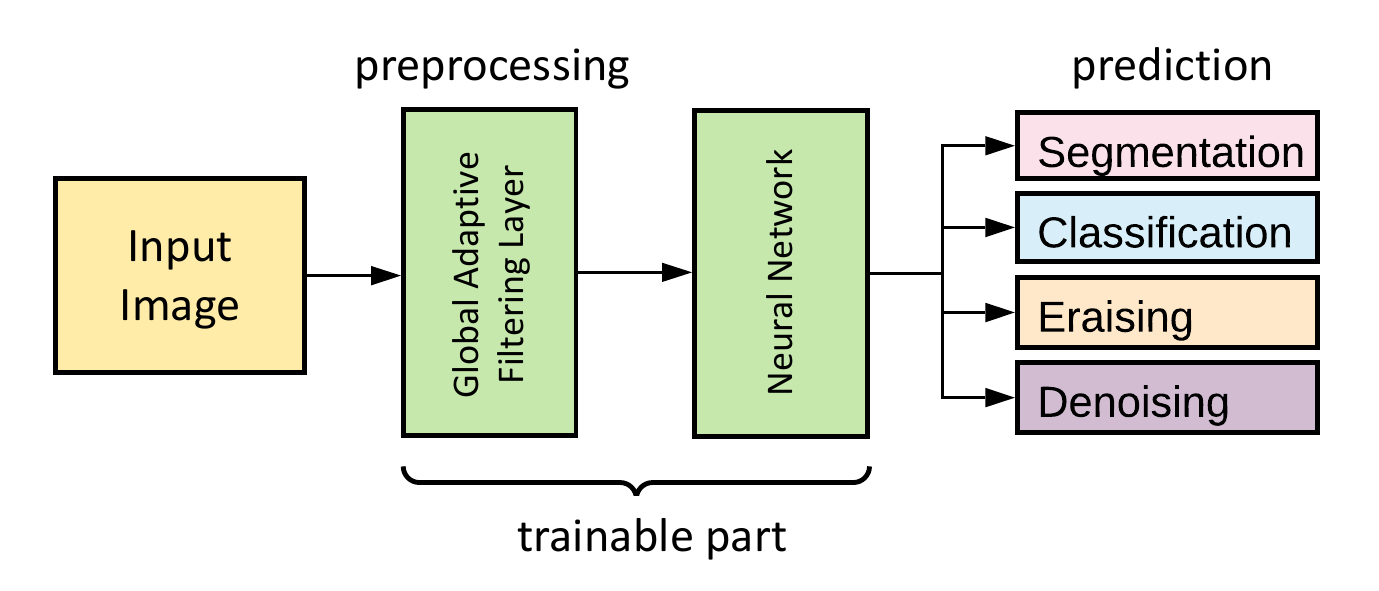}
\caption{Diagram of the proposed method. Global Adaptive Filtering Layer is trained together with the weights of the main neural network until the prediction is maximized in a given CV task.}
\label{fig:gafl}
\end{center}
\end{figure}

Data preprocessing is an essential part of any CV algorithm \cite{szeliski2011computer}, primarily done in the image space \cite{Buslaev_2020_Albumentations}.
Proposing the same in the Fourier space, we want to dismiss the meaningless features associated with spectral frequencies brought to the scene by the image acquisition systems (an ultrasound machine or a photo camera, in our case). As such, the method we look for belongs to the class of \textit{minimum features inductive bias} algorithms \cite{Gordon1995InductiveBias}. The desired `smart' spectral preprocessing method should automatically distill the meaningful frequencies, being aware of the entire model and `adapting' to the entire dataset.

We propose the concept of such a globally adaptive neural layer (see Fig. \ref{fig:gafl}), which could be trained together with a model of interest to solve a given CV problem. By placing this layer in front of the baseline model, the algorithm should automatically select the weights for the frequency components of all images sent in as the input to carry out filtering with \textbf{one purpose only}: \textit{improve whatever the main architecture attempts to do}.

Theoretically \cite{Blackledge,Klette,Brunton}, while performing the Fourier transform, it is possible to move from the frequency domain to the spatial one (and vice versa) without a loss of information. 
The Fourier transform has the property of linearity, preserving the accuracy of signal approximation (Parseval's theorem~\cite{Blackledge}), which is valid when the signal is represented using discrete vectors. 
Define the Fourier transform as
\[
\mathcal{F} I(\uu, \vv) = \sum\limits_{\xx = 0}^{n - 1}\sum\limits_{\yy = 0}^{m - 1} 
\frac{ I(\xx, \yy)  }{nm} \exp \left\{  - \frac{2\pi i\xx \uu}{n} - \frac{2\pi i \yy \vv}{m} \right\} ,
\]
where $I$ is the original image (spatial description) of size $n \times m$, and $\mathcal{F} I$ is the frequency domain image. If the image has multiple channels, we transform each channel separately by the same formula.

For the visual analysis of Fourier transform, one usually works with the spectrum\footnote{We do not center our discussion around \textit{phase}, which could also prove useful for some applications. See Supplementary material.}, \textit{i.e.}, the coordinate-wise absolute value $ | \mathcal{F} I| $, or the energy spectrum $|\mathcal{F} I |^2$. 
To filter image in the frequency domain, we choose to take a function that modifies spectrum $ | \mathcal{F} I| $ in a specific way. 
There is flexibility in designing filtering functions~\cite{Klette}. 
\textit{E.g.}, one can independently select the frequencies to suppress or enhance; however, due to the wide variety of options and the specifics of each task, it is very difficult to select them manually.  
In contrast, the proposed design of the filtering layer shown in Fig.~\ref{fig:gafl} is capable of automatically forming a more sophisticated and a task-specific filter
\footnote{Basic Fast Fourier Transform (FFT) and the element-wise multiplication functions in modern software packages are suitable.}.


To approximate an arbitrary nonlinear function that performs the desired filtering, a neural network with just one hidden layer is enough, yielding:
\[ 
|\mathcal{F} I| \gets W_2 * \sigma{\Bigl(W_1 * |\mathcal{F} I| + B_1\Bigr)} + B_2,
\]
where 
$W_1$ and $W_2$ are the weight matrices with non-negative elements, 
$B_1$ are $B_2$ are the bias matrices with non-negative elements, 
$\sigma(\cdot)$ is some nonlinear activation function,
and ``$*$'' denotes element-wise multiplication.
Algorithm~\ref{alg:Fourier2dLayer} describes the complete function of the proposed global adaptive filtering layer.
\begin{algorithm}[b]
   \caption{Global Adaptive Filtering Layer}
   \label{alg:Fourier2dLayer}
\begin{algorithmic}[1]
    \medskip
\STATEx {\textbf{Input:} $I$ -- Initial image.}
    \STATEx \quad\qquad$\mathcal{F}$ -- Fast Fourier Transform operator.
    \medskip
    \STATE \textit{$W_1$}, \textit{$W_2$}, \textit{$B_1$}, \textit{$B_2$} = \textit{ReLU}(
    $W_1$, $W_2$, $B_1$, $B_2$)\;
    \STATE $F$ = $\mathcal{F} I$\;
    \STATE $S$ = $W_2 * \sigma{\Bigl(W_1 * |F| + B_1\Bigr)} + B_2$\;
    \STATE $S$ = $S * F / |F|$\;
    \STATE $I'$ = $\mathcal{F}^{-1} S $\;
    \medskip
\STATEx {\textbf{Output:} $I'$ -- Image after global frequency filtering.}
\end{algorithmic}
\end{algorithm}

\paragraph{Handling small frequency values.} When a base neural network is chosen, our layer is pre-pended to it and, then, evaluated with several variations to experiment with the small values of non-central frequencies.
Namely, for the \emph{Linear} configuration, a simple single layer neural network of Algorithm \ref{alg:Fourier2dLayer} is used. 
For the \emph{General} configuration, the exponential transform was additionally applied:
$$|\mathcal{F} I| \gets \exp{\Bigl[W * \log{(1 + |\mathcal{F} I|)}\Bigr]} - 1,$$
where 
$W$ is the trained weight matrix (non-negative).
For the \emph{Linear log} and the \emph{General log} configurations, only the log transforms of the spectra were computed.
These configurations are designed to `boost' the appearance of the smallest frequency pixel values in the spectrum, where the intensity of a typical central frequency oftentimes `overwhelms' the smaller values on the periphery (see insets in Fig.~\ref{fig:segmentation_results} below to see typical spike-shaped learnt spectra).

\paragraph{Number of parameters.} The complex values of Fourier transform are tackled by the operation $|\mathcal{F}I|$. Yet, the important symmetry property $\mathcal{F}I(n - u, m - v) = \mathcal{F}I(-u, -v) = \mathcal{F}I(u, v)^{*}$ helps to compute the number of parameters added by the adaptive layers. Namely, each matrix of weights is a tensor of size $\bigl(C, n, \lfloor{m/2\rfloor} + 1\bigr)$, where $C$ is the number of channels, $(n, m)$ is the image size.
Thus, the number of learnable parameters of the proposed adaptive layer is equal to the number of weight matrices multiplied by the product of the matrix dimensions (as the operations in the frequency space are element-wise).


\section{Experiments and Results}
\label{section:experiments}

In this section, we compare the performance of renowned neural networks \textit{with and without} the proposed trainable layers. In all four tasks, we always choose the most popular models, common initialization strategies, and only the well-known activation and loss functions. We aim to make the existing network architectures more efficient.

\noindent\textbf{Hyperparameters common to all tasks.} All models are trained setting the input size to (256, 256), batch size 4, and using \emph{Adam optimizer}~\cite{Kingma} with learning rate 0.001.

\subsection{Datasets}
\label{section:datasets}

We validate efficiency of our adaptive layer on two medical and on four natural image benchmarks. 

The medical benchmarks comprise ultrasonic datasets, popular in medical vision community: \textit{Breast Ultrasound Images} \big(\textbf{BUSI}~\cite{Dhabyani}, 1578 images of three classes: normal (266), benign (891) and malignant (421) tumors, as well as ground-truth segmentation masks\big) and \textit{Brachial Plexus Ultrasound Images} \big(\textbf{BPUI}~\cite{BPUI_dataset}, 5635 images and masks\big).

The natural benchmarks were selected to represent well-known datasets of various scales: from the small \textbf{Caltech Birds} \big(Caltech-UCSD Birds-200-2011~\cite{WahCUB_200_2011}, 11,788 images of 200 classes with ground-truth segmentation masks\big), to medium \textbf{Dogs \textit{vs.}~Cats} \big(\cite{dogs_vs_cats}, 25k images for binary classification\big), to large-scale \textbf{CIFAR-10} \big(\cite{Krizhevsky}, 50k images of 10 classes\big) and a part of \textbf{ImageNet} \big(`Tiny' ImageNet \cite{Le}, 110k images of 200 classes\big).

We considered medical imaging datasets separately because the ultrasound signal is known to have particular frequency bands needed for the optimal image contrast in live imaging~\cite{Ihnatsenka}, making us hypothesize that the filtering effect would be more pronounced in these data. Dataset details and descriptions are given in the Supplementary material.

\begin{figure}[t]
\begin{center}
\includegraphics[width=\columnwidth]{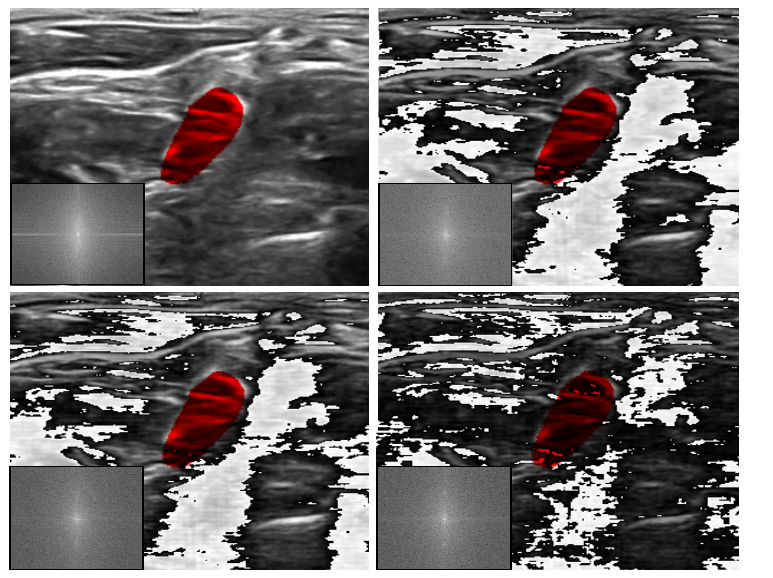}
\caption{\textsc{Segmentation Results}. Examples of 'learnt' filters and their effect on the segmentation. Top left: original image and ground-truth mask; top right: \emph{Linear} filter; bottom left: \emph{General} filter; bottom right: \emph{General log} filter. Corresponding `learnt' spectra are shown in the insets in the corner.}
\label{fig:segmentation_results}
\end{center}
\end{figure}

\subsection{Segmentation}
To test the proposed method, three different networks were studied as the base models: U-Net~\cite{Ronneberger}, DenseNet~\cite{Densenet}, and ResNet~\cite{Resnet}.
%
\begin{figure}[b]
\begin{center}
\includegraphics[width=\columnwidth]{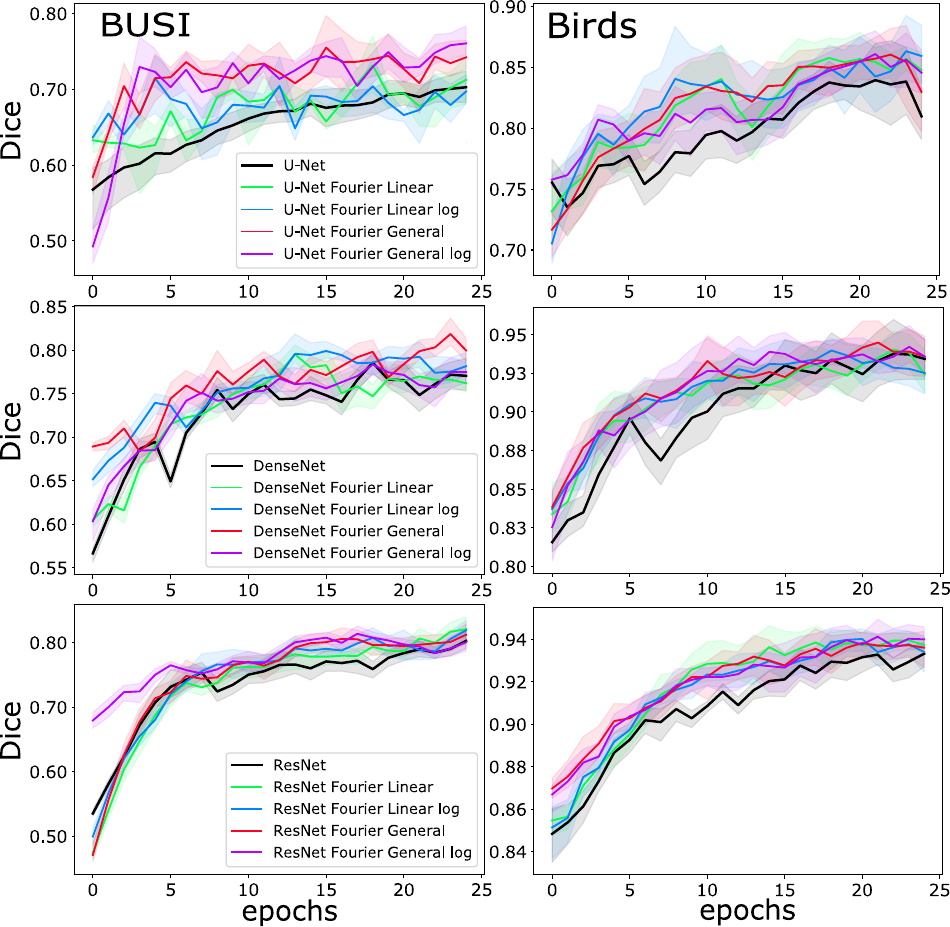}
\caption{\textsc{Segmentation Results}. Average Dice coefficients on validation sets of different datasets: medical (BUSI) and natural (Birds). Top row:  U-Net, middle: DenseNet, bottom: ResNet.}
\label{fig:segmentation_results_metrics}
\end{center}
\end{figure}
For the learning process, we used the Combined Loss function of \emph{Dice} and \emph{Cross Entropy}, weighted as 0.6 and 0.4 respectively. 
The quality of segmentation is evaluated with the \emph{Dice coefficient}~\cite{Milletari}, which, in essence, measures the overlap between the predicted and the ground-truth masks.

The following hyperparameters were used. For U-Net, \emph{init\_features} = 32 (number of parameters in initial convolution), \emph{depth} = 3 (number of downsteps). 
For DenseNet (as for densenet-121), \emph{init\_features} = 32, \emph{growth\_rate} = 32 (number of filters to add to each layer), \emph{block\_config} = 6, 12, 24, 16 (number of layers in each pooling block).
For ResNet (as for resnet-18), \emph{blocks}: 2, 2, 2, 2 (number of layers in each pooling block). 
\begin{table}[!ht]
\caption{\textsc{Segmentation Results}. Validation Dice score for different models on medical (BUSI and BPUI) and natural (Caltech Birds) datasets. Red: the worst model, green: the best model.}
\label{tab:segmentation_val_metrics}
\begin{center}
\begin{small}
\begin{tabular*}{\columnwidth}{@{\extracolsep{\fill}}lcccccc}
\toprule
\textbf{Model} & \textbf{BUSI} & \textbf{BPUI} & \textbf{Birds} \\
\bottomrule
\textcolor{red}{{U-Net}} & \textcolor{red}{{0.70}} & \textcolor{red}{{0.59}} & \textcolor{red}{{0.84}} \\
\midrule
+ Fourier Linear & 0.72 & 0.64 & 0.85 \\
\midrule
+ Fourier Linear log & 0.71 & 0.64 & \textcolor{dgreen}{{0.86}} \\
\midrule
\textcolor{dgreen}{{+ Fourier General}} & \textcolor{dgreen}{{0.75}} & \textcolor{dgreen}{{0.74}} & \textcolor{dgreen}{{0.86}} \\
\midrule
\textcolor{dgreen}{{+ Fourier General log}} & \textcolor{dgreen}{{0.75}} & \textcolor{dgreen}{{0.74}} & \textcolor{dgreen}{{0.86}} \\
\bottomrule
\textcolor{red}{{DenseNet}} & \textcolor{red}{{0.77}} & \textcolor{red}{{0.74}} & \textcolor{red}{{0.94}} \\
\midrule
+ Fourier Linear & 0.79 & \textcolor{dgreen}{{0.77}}  & \textcolor{red}{{0.94}} \\
\midrule
+ Fourier Linear log & 0.80 & \textcolor{dgreen}{{0.77}} & \textcolor{red}{{0.94}} \\
\midrule
\textcolor{dgreen}{{+ Fourier General}} & \textcolor{dgreen}{{0.81}} & 0.74  & \textcolor{dgreen}{{0.95}} \\
\midrule
+ Fourier General log & \textcolor{red}{{0.77}} & 0.75 & \textcolor{red}{{0.94}} \\
\bottomrule
\textcolor{red}{{ResNet}} & \textcolor{red}{{0.80}} &  \textcolor{red}{{0.71}} & \textcolor{red}{{0.93}} \\
\midrule
+ Fourier Linear  & \textcolor{dgreen}{{0.81}} & 0.72 & \textcolor{dgreen}{{0.94}} \\
\midrule
+ Fourier Linear Log & \textcolor{dgreen}{{0.81}} & 0.72 & \textcolor{dgreen}{{0.94}} \\
\midrule
+ Fourier General & \textcolor{dgreen}{{0.81}} & 0.74 &  \textcolor{dgreen}{{0.94}} \\
\midrule
\textcolor{dgreen}{+ Fourier General Log} & \textcolor{dgreen}{{0.81}} & \textcolor{dgreen}{{0.75}} &  \textcolor{dgreen}{{0.94}} \\
\bottomrule
\end{tabular*}
\end{small}
\end{center}
\end{table}

We observe improvement of the segmentation performance in all three base models, as summarized in Figs.~\ref{fig:segmentation_results},~\ref{fig:segmentation_results_metrics}, and in Table~\ref{tab:segmentation_val_metrics}. Comprehensive results for each dataset and each model are given in the Supplementary material. 
%
\subsection{Classification}

\begin{figure}[!ht]
\begin{center}
\includegraphics[width=\columnwidth]{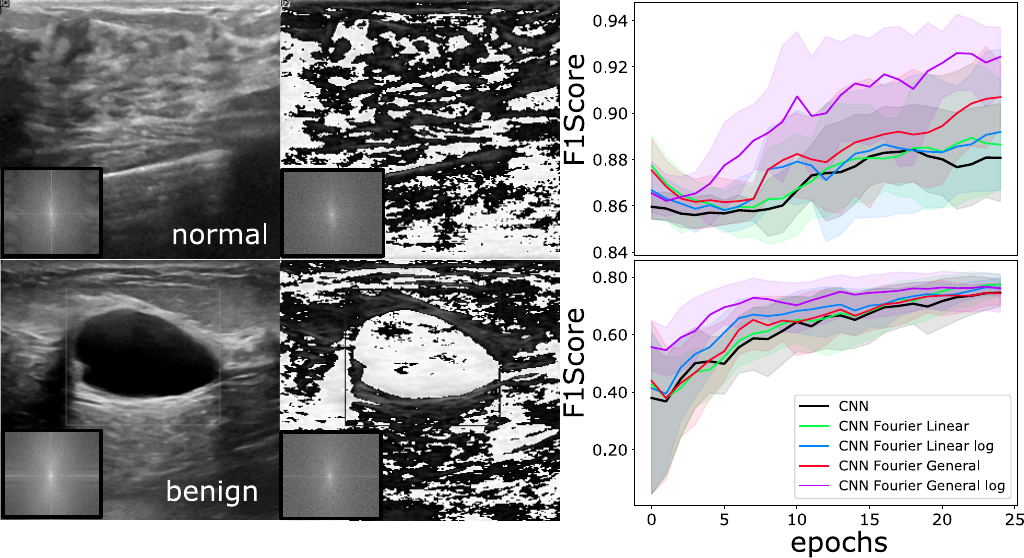}
\caption{
\textsc{Classification Results: Binary}.
Left: Initial and filtered images using our layer in \emph{General log} configuration. Insets show the `learnt' optimal spectra.
 Right: $F_1$-scores on BUSI validation sets (normal \textit{vs.} benign and benign \textit{vs.} malignant). 
}
\label{fig:classification_results_BUSI}
\end{center}
\end{figure}

To verify the suggested algorithm for the classification problem, a typical Convolutional Neural Network (CNN) with several convolutional blocks and fully-connected layers is used. 
Namely, the encoder blocks include \emph{conv}, \emph{Batch Normalization}, \emph{ReLU}, \emph{Average Pooling}, and two \textit{fully-connected} layers (using \emph{init\_features} = 8 and \emph{depth} = 4).
The training process is similar to the one above, with using the \emph{weighted Cross Entropy Loss}~\cite{Ho} combined with the \emph{$F_1$-score} evaluation. 
The results for this task for medical and natural datasets are presented in Fig.~\ref{fig:classification_results_BUSI} and Table~\ref{tab:classification_val_metrics}. Additional natural image classification experiments were performed on large-scale datasets CIFAR-10 and ImageNet, with the results presented in Table \ref{tab:control_experiments_val_metrics} and Fig. \ref{fig:control_experiments_results}.

\begin{table}[!ht]
\caption{\textsc{Classification Results: Binary}. $F_1$-scores for different models on BUSI validation set (normal \textit{vs.} benign and benign \textit{vs.} malignant classes), and Dogs \textit{vs.} Cats datasets. Red: the worst model, green: the best model.}
\label{tab:classification_val_metrics}
\begin{center}
\begin{small}
\begin{tabular*}{\columnwidth}{@{\extracolsep{\fill}}lcccccc}
\toprule
\multirow{2}{*}{\textbf{Model}} & \multicolumn{1}{c}{\textbf{BUSI}} & \multicolumn{1}{c}{\textbf{BUSI}} & \multicolumn{1}{c}{\textbf{D \textit{vs.} C}} \\
& \multicolumn{1}{c}{norm \textit{vs.} ben} & \multicolumn{1}{c}{ben \textit{vs.} mal} & \multicolumn{1}{c}{} \\
\bottomrule
\textcolor{red}{{CNN}} & \textcolor{red}{{0.88}} & \textcolor{red}{{0.76}} & \textcolor{red}{{0.82}} \\
\midrule
\textcolor{dgreen}{+ Fourier Linear} & 0.89 & \textcolor{dgreen}{{0.78}} & \textcolor{dgreen}{{0.83}} \\
\midrule
+ Fourier Linear log & 0.89 & 0.77 & \textcolor{red}{{0.82}} \\
\midrule
+ Fourier General & 0.90 & \textcolor{red}{{0.76}} & \textcolor{red}{{0.82}} \\
\midrule
\textcolor{dgreen}{+ Fourier General log} & \textcolor{dgreen}{{0.92}} & 0.77 & \textcolor{dgreen}{{0.83}} \\
\bottomrule
\end{tabular*}
\end{small}
\end{center}
\end{table}


\begin{figure*}[!ht]
\begin{center}
\includegraphics[width=\textwidth]{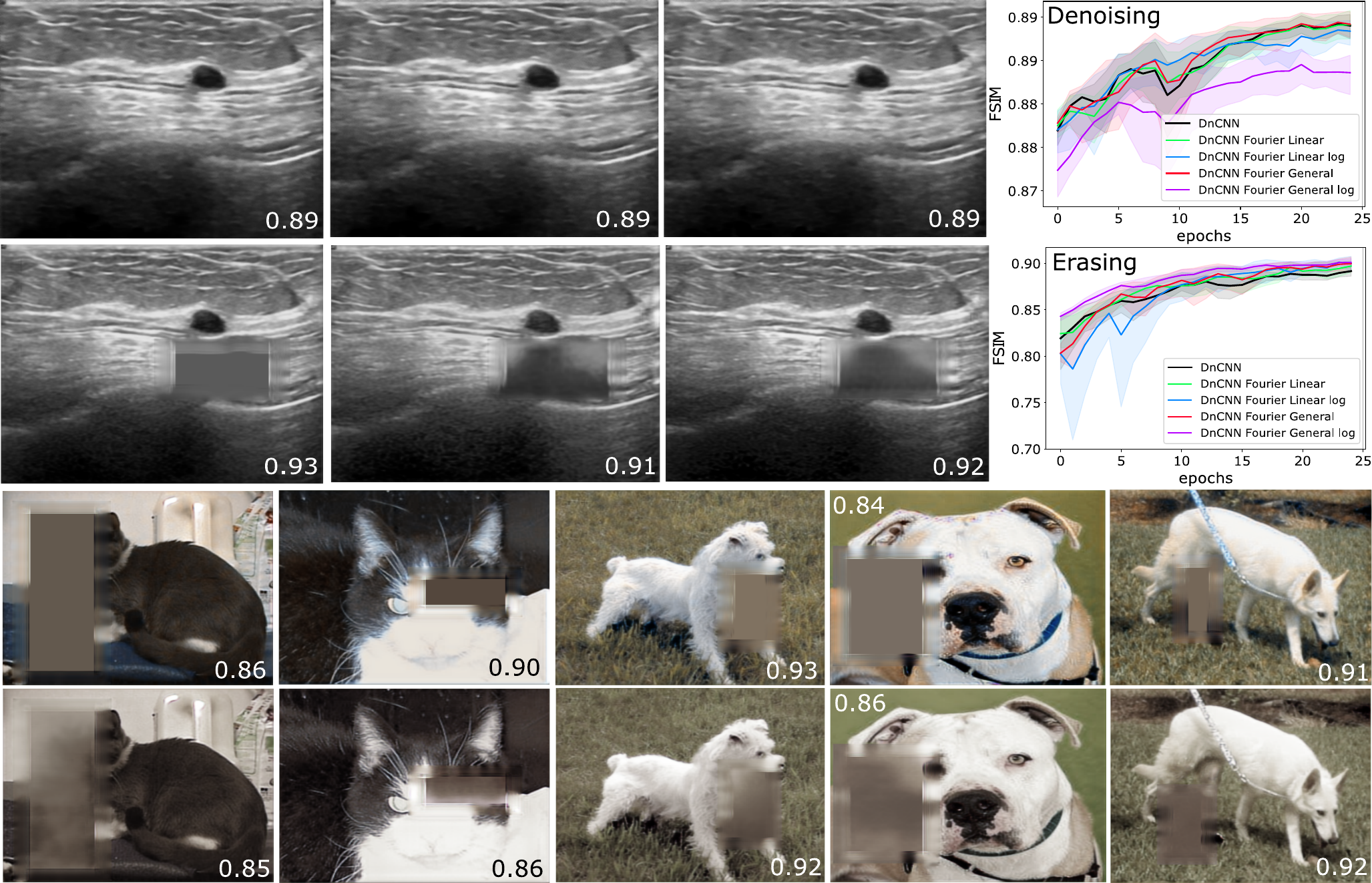}
\caption{\textsc{Denoising/Erasing Results}. Plots on the right show FSIM metrics for denoising and erasing corruptions of BUSI dataset for different models. First and second row images, left to right: base DnCNN model results, the model with \emph{Linear}, and with \emph{General} adaptive filters on BUSI dataset. Third and fourth rows, left to right: base model and model with \emph{Linear} adaptive filtering layer on the Dogs \textit{vs.} Cats dataset. Note how addition of Fourier-based layer corrects for the corruptions better (\textit{e.g.}, dog images in fourth \textit{vs.} third row).}
\label{fig:panel_denoising}
\end{center}
\end{figure*}

\subsection{Denoising and Erasing}

For the problem of denoising and erasing, we considered popular model DnCNN~\cite{Zhang} as the baseline, the main task of which is to restore the noise, in contrast to the standard feedforward models, which restore the image.
To assess the success of denoising and erasing, we used Combined Loss function of \emph{MS-SSIM} and \emph{$L_1$ Loss} with weights 0.8 and 0.2 respectively~\cite{Zhao} along with the \emph{FSIM} and \textit{PSNR} metrics~\cite{Zhang_FSIM}. 
Regular Gaussian noise and rectangular erasing corruptions~\cite{Zhong} were introduced to the images, with the consequent image recovery yielding the outcomes summarized in Fig.~\ref{fig:panel_denoising}.
For the denoising and erasing problems, we used the following hyperparameters: \emph{init\_features} = 64, \emph{num\_layers} = 17 (number of layers).

\subsection{Control and large-scale experiments} 
\begin{figure}[!ht]
\begin{center}
\includegraphics[width=\columnwidth]{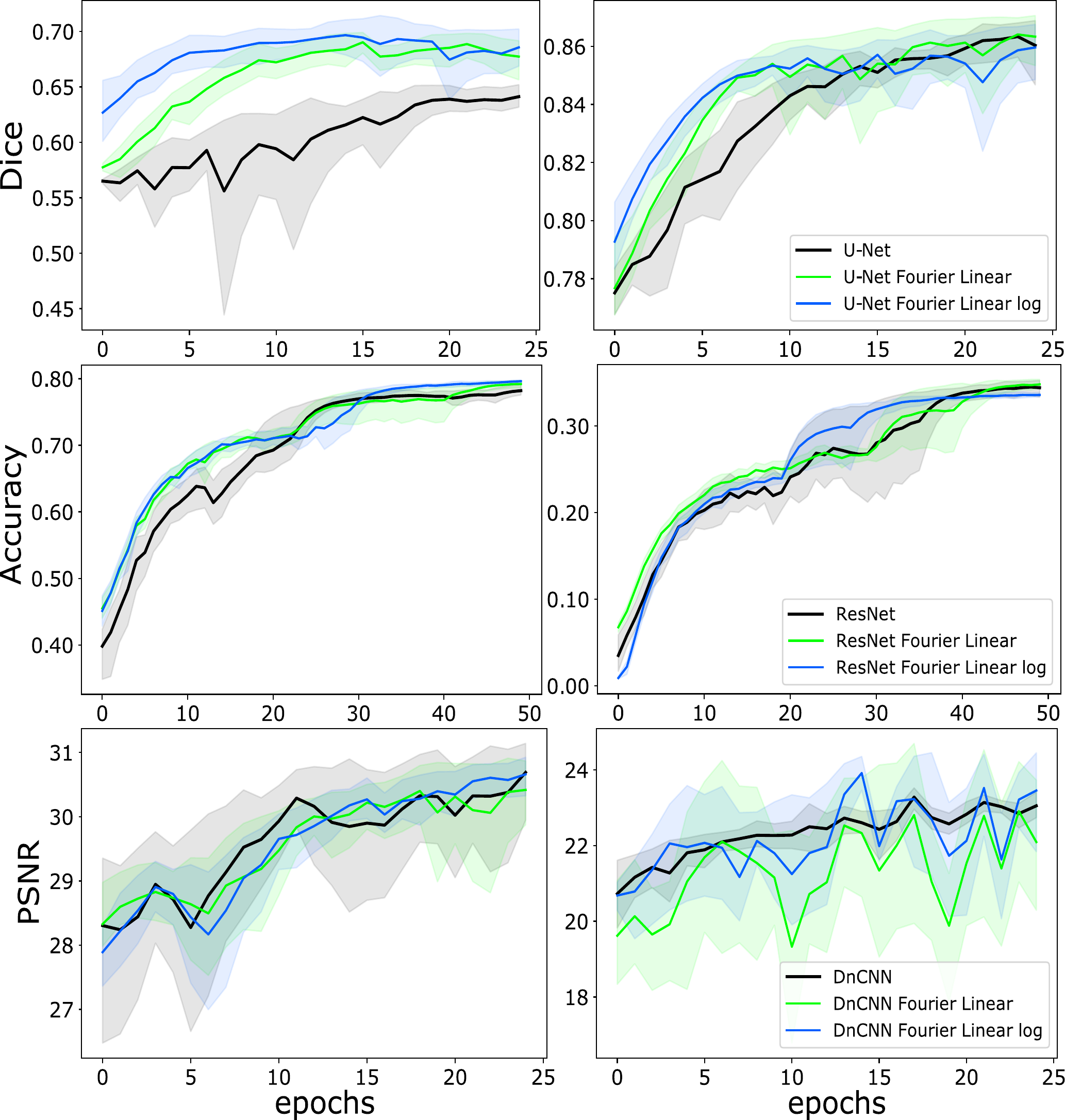}
\caption{\textsc{Control and Large-Scale Experiments}. Metrics on validation sets for different models. 
Top row: Dice score for segmentation problem on medical (BUSI, left) and natural (Caltech Birds, right) datasets. 
Middle row: Prediction accuracy for classification problem on CIFAR-10 (left) and Tiny ImageNet (right) datasets. 
Bottom row: PSNR metric for Gaussian denoising (left) and erasing corruption (right) problems on BUSI dataset.}
\label{fig:control_experiments_results}
\end{center}
\end{figure}

To gauge the impact of the added layer on the base model \textit{precisely}, a fine control of the total number of trainable parameters is desirable. 
Therefore, additional studies were performed for each CV problem, assuring that the number of parameters in base models was \textit{either greater or equal} of that with the pre-pended adaptive filtering layer (Table~\ref{tab:control_experiments_val_metrics} and Fig.~\ref{fig:control_experiments_results}). 
In this sub-study, all experiments were run using the \textit{Linear} configurations of the adaptive layer. 
\begin{table*}[t]
\caption{\textsc{Control Experiments Results}. Comparison between different models for segmentation (Dice score), classification (Accuracy), and Gaussian denoising and erasing corruptions (PSNR). The number of parameters in all experiments is controlled not to exceed the number of parameters in the corresponding base models.}
\label{tab:control_experiments_val_metrics}
\begin{center}
\begin{small}
\begin{tabular*}{\textwidth}{@{\extracolsep{\fill}}lcccccc}
\toprule
\multirow{2}{*}{\textbf{Model}} & \multicolumn{2}{c}{\textbf{Segmentation}} & \multicolumn{2}{c}{\textbf{Classification}} & \multicolumn{1}{c}{\textbf{Denoising}} & \multicolumn{1}{c}{\textbf{Erasing}} \\
& \multicolumn{1}{c}{BUSI} & \multicolumn{1}{c}{Birds} & \multicolumn{1}{c}{CIFAR-10} & \multicolumn{1}{c}{Tiny ImageNet} & \multicolumn{1}{c}{BUSI} & \multicolumn{1}{c}{BUSI} \\
\bottomrule
\textcolor{red}{Base model} & \textcolor{red}{0.64} & \textcolor{dgreen}{0.86} & \textcolor{red}{0.782} & 0.345 & \textcolor{dgreen}{30.69} & 23.28\\
\midrule
+ Fourier Linear & 0.69 & \textcolor{dgreen}{0.86} & 0.792 & \textcolor{dgreen}{0.348} & \textcolor{red}{30.42} & \textcolor{red}{22.91} \\
\midrule
\textcolor{dgreen}{+ Fourier Linear log} & \textcolor{dgreen}{0.70} & \textcolor{dgreen}{0.86} & \textcolor{dgreen}{0.796} & \textcolor{red}{0.336} & 30.66 & \textcolor{dgreen}{23.92}\\
\bottomrule
\end{tabular*}
\end{small}
\end{center}
\end{table*}

For the segmentation task, U-Net with \textit{init\_features} = 32 was compared against U-Net with \textit{init\_features} = 16 along with adaptive layer for BUSI (with \textit{image\_size} = (512, 512), yielding 467,266 parameters in base model \textit{vs.} 248,994 parameters in ours) and Caltech Birds (with \textit{image\_size} = (256, 256), yielding 467,842 parameters in base model \textit{vs.} 216,770 parameters in ours) datasets.

For the classification task, ResNet-20 was compared with the same model pre-pended by the adaptive layer for CIFAR-10 (with \textit{image\_size} = (32, 32), yielding 276,026 \textit{vs.} 274,394 parameters) and Tiny ImageNet (with \textit{image\_size} = (64, 64), yielding 293,080 \textit{vs.} 286,744 parameters) datasets. For these experiments, \textit{SGD optimizer} with weight decay of 0.0001, momentum of 0.9, and learning rate 0.1 was used.

For the denoising and erasing tasks, DnCNN with \textit{init\_features} = 32 and \textit{num\_layers} = 20 was compared to itself with \textit{init\_features} = 16 and \textit{num\_layers} = 17 and with the pre-pended adaptive layer for BUSI (with \textit{image\_size} = (512, 512), yielding 168,225 \textit{vs.} 167,169 parameters).

\section{Discussion}
\label{section:discussions}

Remarkably, in the controlled experiments, we observed that the number of parameters in the base model could be reduced by half; yet, the global adaptive filtering layer allows to `catch up' with the lost parameters and to reach the level of the base models that have twice as many parameters. The result generalize well across different datasets of various scales and across CV tasks. We did not observe a stunning improvement in the denoising problem built around DnCNN model -- a result we attribute to the way the noise is handled in the Fourier domain, making our metric choice somewhat sub-optimal. Additional studies are required with the DnCNN to understand the enhancement dynamics; however, the same very model is well improved by our layer when there is a notable corruption (the erasing problem). We can visually confirm that our adaptive configuration denoises and `heals' the corruptions better than the base DnCNN model alone.

\paragraph{Activation Functions.} The activation function used in the general configuration of proposed global filtering layer is a hyperparameter that needs to be selected depending on the problem being solved and the dataset. In the provided algorithm, the function receives a non-negative matrix as input; it is since the frequency with negative weight has no physical interpretation. Therefore, several popular activation functions have been selected and investigated. As can be seen in the results in the Supplementary material, the activation function plays an important role. The average difference between the best and worst activation functions in some cases can lead to about 10\% gain by metric. It should be noted that the activation functions \emph{Mish} and \emph{ReLU} have shown themselves well on all datasets. So, \emph{Mish} and \emph{ReLU} are good for using our algorithm out of the box. Note that all the experiments reported in the main text were carried out using \emph{ReLU} activation function.

\paragraph{Medical vs. Natural Datasets.} In datasets collected using ultrasound imaging, there are a lot of negative examples (with a zero or an empty mask). Unlike the Caltech Birds dataset, such data require more focus, dedication, and expertise to annotate them with labels. But, as with all human-tagged data, one should expect artefacts and potential errors in the markings: for example, the BPUI data have been annotated by pseudo-experts (people who have been trained and instructed by experts).

\begin{figure*}[]
\begin{center}
\includegraphics[width=\textwidth]{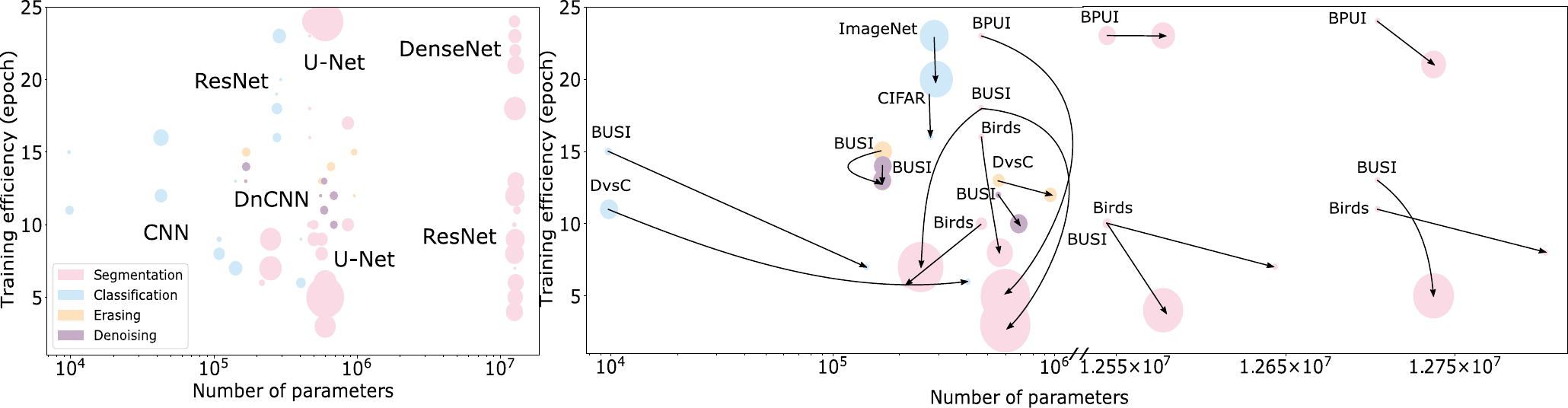}
\caption{Training efficiency as a function of model complexity for all 88 experiments. The bubble size encodes \textbf{the gain in the corresponding metrics}. Left: all experiments; right: zoomed-in areas. Colours correspond to different types of 4 CV tasks considered. Vertical and left-leaning arrows correspond to 18 \textbf{control experiments} run with precise control of the number of parameters in the models. The arrows indicate the correspondence of pairs (basic model $\to$ the most training-efficient model with the proposed adaptive Fourier layer). Birds: Caltech Birds (2011) dataset; DvsC: Dogs \textit{vs.} Cats dataset; CIFAR: CIFAR-10 dataset; ImageNet: Tiny ImageNet dataset.}
\label{fig:panel_bubble_chart}
\end{center}
\end{figure*}

Several important observations and conclusions can be made from the results in the figures herein and from those in the Supplementary material:

\begin{itemize}[leftmargin=*]
    \item The observed improvement of Dice score on the ultrasound datasets depends on the number of parameters of the base model: the larger the architecture size, the less noticeable the increase in the metric when using the adaptive layer.
    \item In the natural images dataset, there is a faster convergence of all models when the proposed filtering layer is added, than in any of the base models.
    \item The proposed log operation applied to the spectra allows for the small pixel values and eliminates fluctuation artefacts otherwise appearing from the truncation.
    \item Addition of the adaptive filter to the natural images provides ``smoother'' convergence of the training curves. We believe this trait could be instrumental for accelerating many state-of-the-art models where predicting the behaviour of the model is important; for example, in reinforcement and/or active learning (Fig.~\ref{fig:panel_bubble_chart}).
\end{itemize}

\paragraph{Erasing metric choice.} At the moment, there are a lot of metrics for assessing the quality of denoising of various types. However, there is no universal one that would accurately correspond to the assessment with the naked eye and is well interpreted in all cases~\cite{Ding}. The most suitable for our task was the FSIM metric, which uses the structure of the Fourier components of the image. However, on a large number of examples (including those provided to you in the Fig.~\ref{fig:panel_denoising}), it was noticed that the difference in metric between the base model and the model with the proposed trainable layer should have a larger gap, since filling the inside of the cropped rectangle has greater importance than just averaging. Our experiments with other metrics, such as PSNR, present a sound but still sub-optimal alternative.


\section{Conclusions}
\label{section:conclusions}

The method proposed in our work proved to be efficient on all datasets and all classical model architectures that we have considered. In all cases, the use of simple adaptive frequency filtering layer has led to faster convergence of the training process than in the case of the stand-alone model, having shown higher segmentation quality both on train and on test samples.
A rather important finding is the increase of Dice score for the case of simple U-Net by around 10\% when the adaptive global filter is added. This promises an opportunity for the areas, such as medicine, where getting marked data is an acknowledged challenge, causing one to attempt learning on small datasets. 
In these cases, the use of heavy models with a large number of parameters is one possible solution which frequently leads to a fast overfit; whereas, addition of simple adaptive filtering layers ``trims'' unnecessary frequencies in the Fourier domain and makes the model learn only the vital frequencies along with the weights of the main neural network. All of this is accomplished while optimizing the targeted advantage function of interest to a particular application (\textit{e.g.} , Dice score or $F_1$-score).

We believe the proposed layer can be a good add-on for a number of powerful modern preprocessing tools, including those that exist in various AutoML pipelines \cite{cubuk2019autoaugment}. In fact, there are many recently devised methods of augmentation and data preprocessing (\textit{e.g.}, see \cite{Buslaev_2020_Albumentations}) that still await for frequency filtering capability. 

The convergence speed of models plays an important role too. Areas, such as ultrasound imaging, entail big amounts of data, the labelling of which takes a lot of time and requires the involvement of highly qualified experts. 
Due to this, in CV is relevant task of active learning~\cite{shelmanov2019active}, which implies the further retraining of models. Same applies to tasks which entail unsupervised segmentation~\cite{bespalov2020brule}.
Initially, we anticipated that our adaptive layer would improve convergence and quality primarily in the ultrasound data (the echogenic nature of which is known to be prone to high sensitivity to the frequency knobs). But we were surprised to find out the improvement in the natural images as well. This expands the possible areas of application of the proposed approach and opens up a new direction of research of adaptive layers (\textit{e.g.} ,in more complex multi-layer architectures~\cite{chowdhury2017blood}, in generative and image translation models~\cite{prokopenko2019unpaired}, in learnable frequency kernels~\cite{lazareva2020learnable}, in iterative anomaly detection models~\cite{tuluptceva2020anomaly}, \textit{etc.}. Lightening of these models via a simple adaptive filtering layer holds potential for a seamless integration in various CV applications.).

\section{Acknowledgements}
We thank Nazar Buzun for helpful discussions.

{\small
\bibliographystyle{main}
\bibliography{main}
}







    

\clearpage
\part*{Supplemental material}
\section{Datasets Description}

\begin{figure}[!ht]
\begin{center}
\includegraphics[width=\columnwidth]{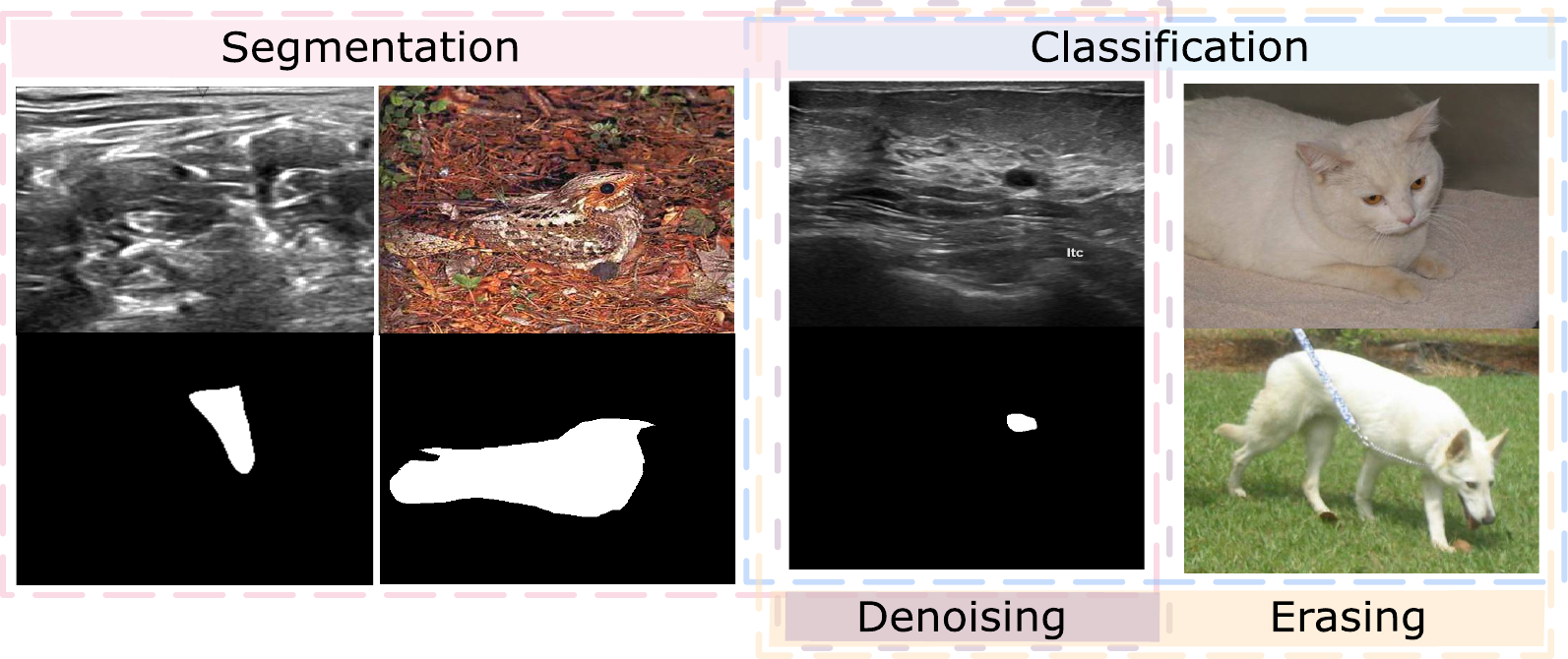}
\caption{Examples of BPUI, Caltech Birds (2011), BUSI, Dogs \textit{vs.} Cats (columns from left to right) datasets associated with the considered computer vision tasks.}
\label{fig:panel_datasets}
\end{center}
\end{figure}

A more exhaustive description of the datasets (Fig.~\ref{fig:panel_datasets}) from the main text, follows below.

We validate efficiency of our adaptive layer on two medical and on four natural image benchmarks.

The ultrasound data is collected by receiving the echo signal reflected from the surfaces of tissues of different density. Depending on the echogenicity (the ability to reflect or transmit ultrasound waves), different types of structures can be distinguished in the scans: hyper-echoic (white regions on the scan), hypo-echoic (gray regions on the scan), or an-echoic (black regions on the scan). 
For example, the muscles are hypo-echoic with a striated structure, and the nerves are hyper-echoic with the stippled structure.

Natural images represent a separate class of images because they are collected by regular cameras.

\paragraph{Breast Ultrasound Images (BUSI) dataset} contains medical ultrasound images of breast cancer for different female patients, along with the ground truth masks. The data was collected to address the problem of classification, detection, and segmentation of breast cancer, one of the most common causes of death in women worldwide. All images are divided into three classes: normal, benign tumor, and malignant tumor.

\paragraph{Brachial Plexus Ultrasound Images (BPUI) dataset} is intended for the task of segmentation of the anatomical networks, formed by the anterior branches of the four lower cervical nerves and the first thoracic nerve, called the brachial plexus. Localization of the nerve structures on the ultrasound scans is an important step in the clinical practice. The dataset contains both the images with the plexus and without, as well as the ground truth masks.

\paragraph{Caltech Birds (CUB\_200\_2011) dataset} provides photographs of 200 different bird species, organized by the classes. Annotations, obtained as the medians over the pixels indicated by 5 different users per image, include the bounding boxes and the segmentation labels. Therefore, for the scope of our work, only the main core contours of the annotation labels are referred to as the ground truth masks.

\paragraph{Dogs \textit{vs.}~Cats dataset} 
consists of a set of 25 thousand images of cats and dogs with a class label. There is a tremendous diversity in the photos (for instance, a wide variety of backgrounds, angles, poses, lighting), causing difficulties for building a proper automatic classification model.

\paragraph{CIFAR-10 dataset} consists of 50,000 images for training and 10,000 images for the test, size $32 \times 32$, which are uniformly distributed over 10 classes: airplane, automobile, bird, cat, deer, dog, frog, horse, ship, truck. Each picture includes only one leading instance of a certain class which can be viewed from an unusual point of view.

\paragraph{Tiny ImageNet dataset} has images of 200 different classes (500 in each) with a size of $64 \times 64$, as well as 50 validation images for each class.

\section{Spectrum and Phase adjustment}

\begin{figure*}[!h]
\begin{center}
\includegraphics[width=0.75\textwidth]{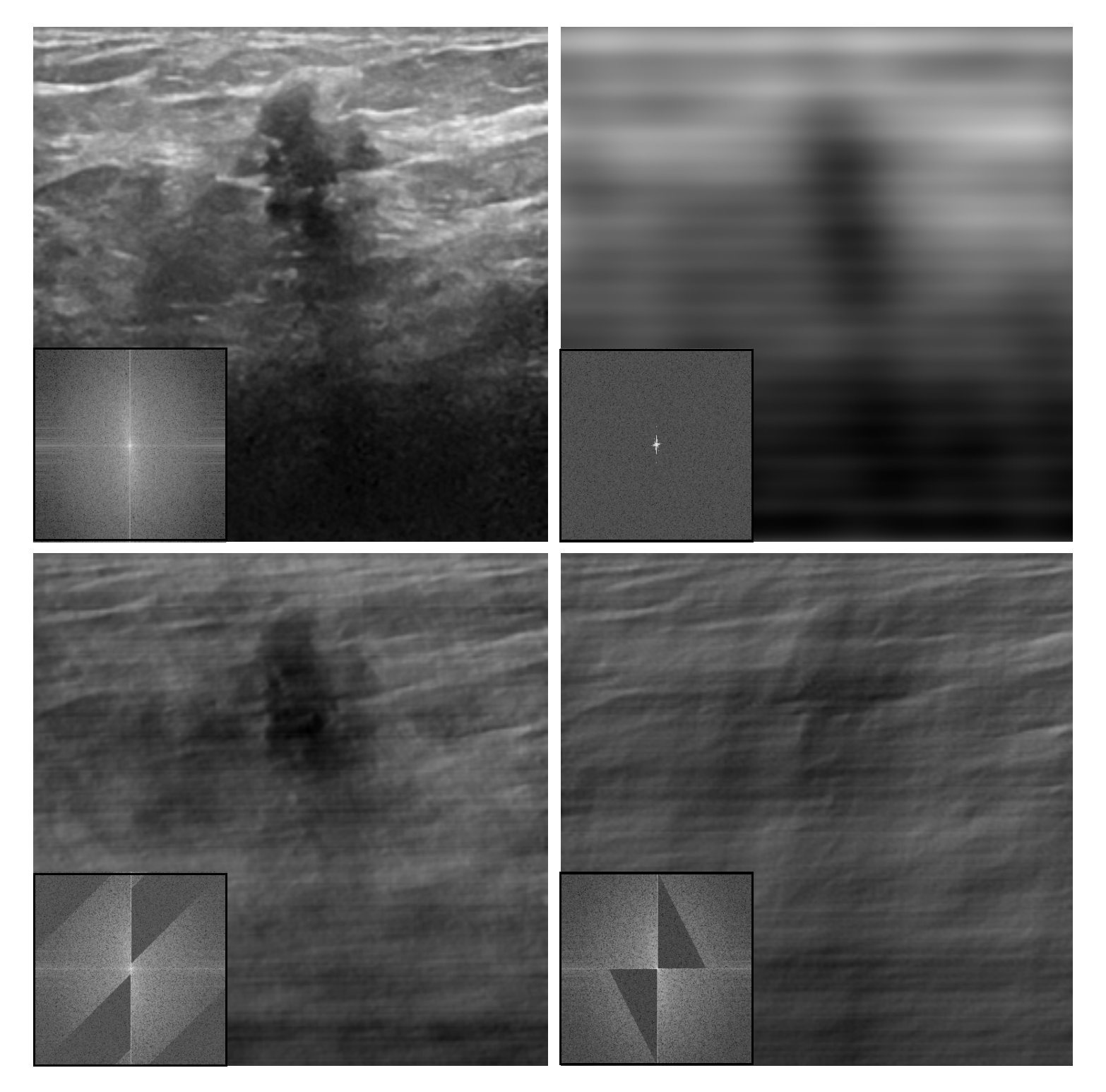}
\caption{Motivation elaborated. Different frequency filters applied to the initial ultrasonic image (top left) and their Fourier spectra. Notice the effect of high-pass/low-pass components on morphology and texture.}
\label{fig:different_filters}
\end{center}
\end{figure*}

\paragraph{Spectrum.} We look for a global filter that will leave only those frequencies that improve the prediction of the base model. A high-pass filter, in which high frequencies are passed and low frequencies are suppressed, enhances edges. While visualizing details with a low-pass filter reduces outliers and contrast, \textit{i.e.}, yielding a smoothing effect. For each task, one needs a specific filter, which is not easy to select manually (Fig.~\ref{fig:different_filters}). Therefore, we can delegate the task of filter selection to the computer by integrating the appropriate adaptive neural layer for global frequency filtering, which will be trained alongside the main model.

\begin{figure*}[!h]
\begin{center}
\includegraphics[width=\textwidth]{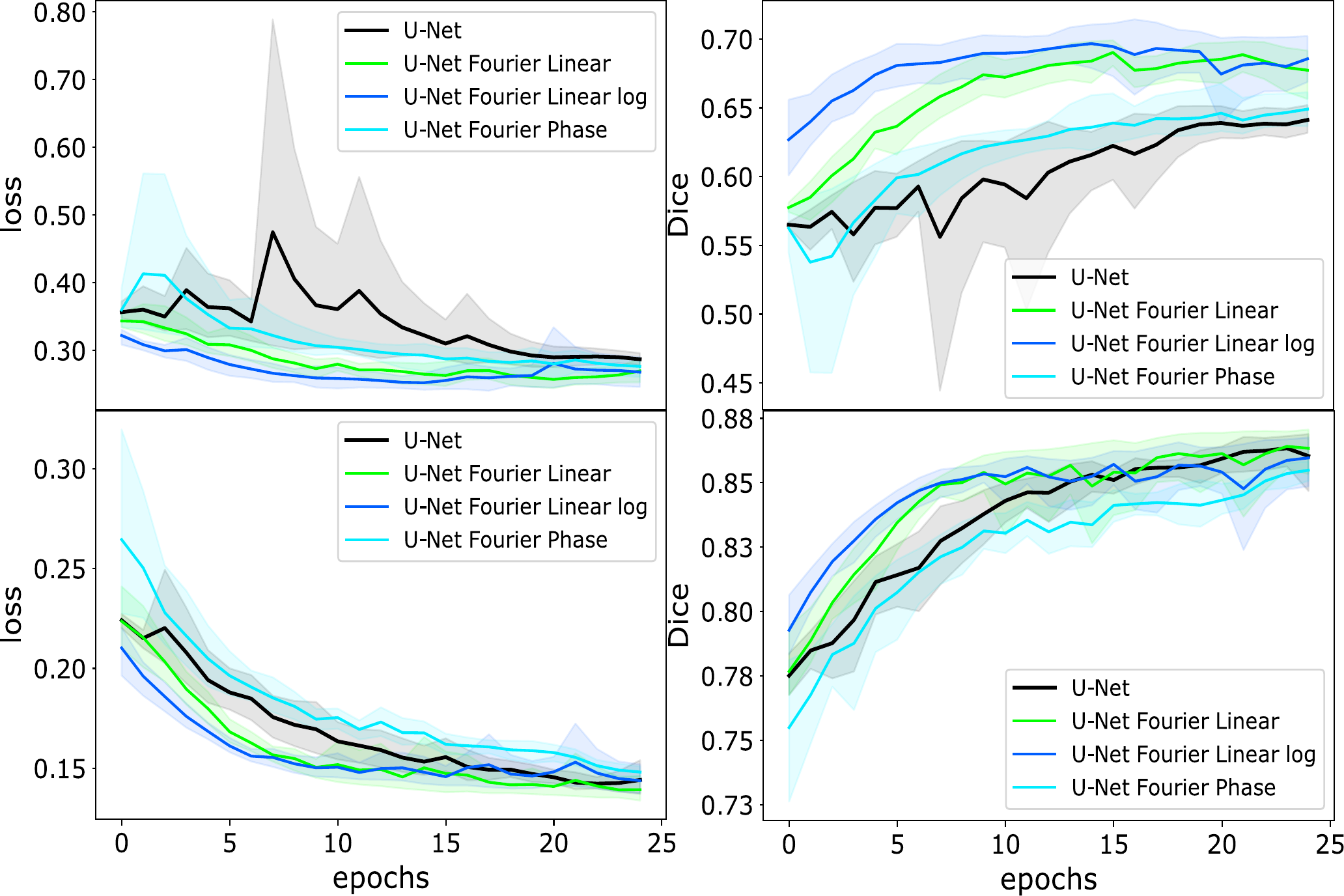}
\caption{\textsc{Control and Large-Scale Experiments}. Loss functions and Dice coefficients of different models training processes on BUSI (top) and Caltech Birds (bottom) \emph{validation} sets.}
\label{fig:control_segmentation}
\end{center}
\end{figure*}

\begin{figure*}[!h]
\begin{center}
\includegraphics[width=\textwidth]{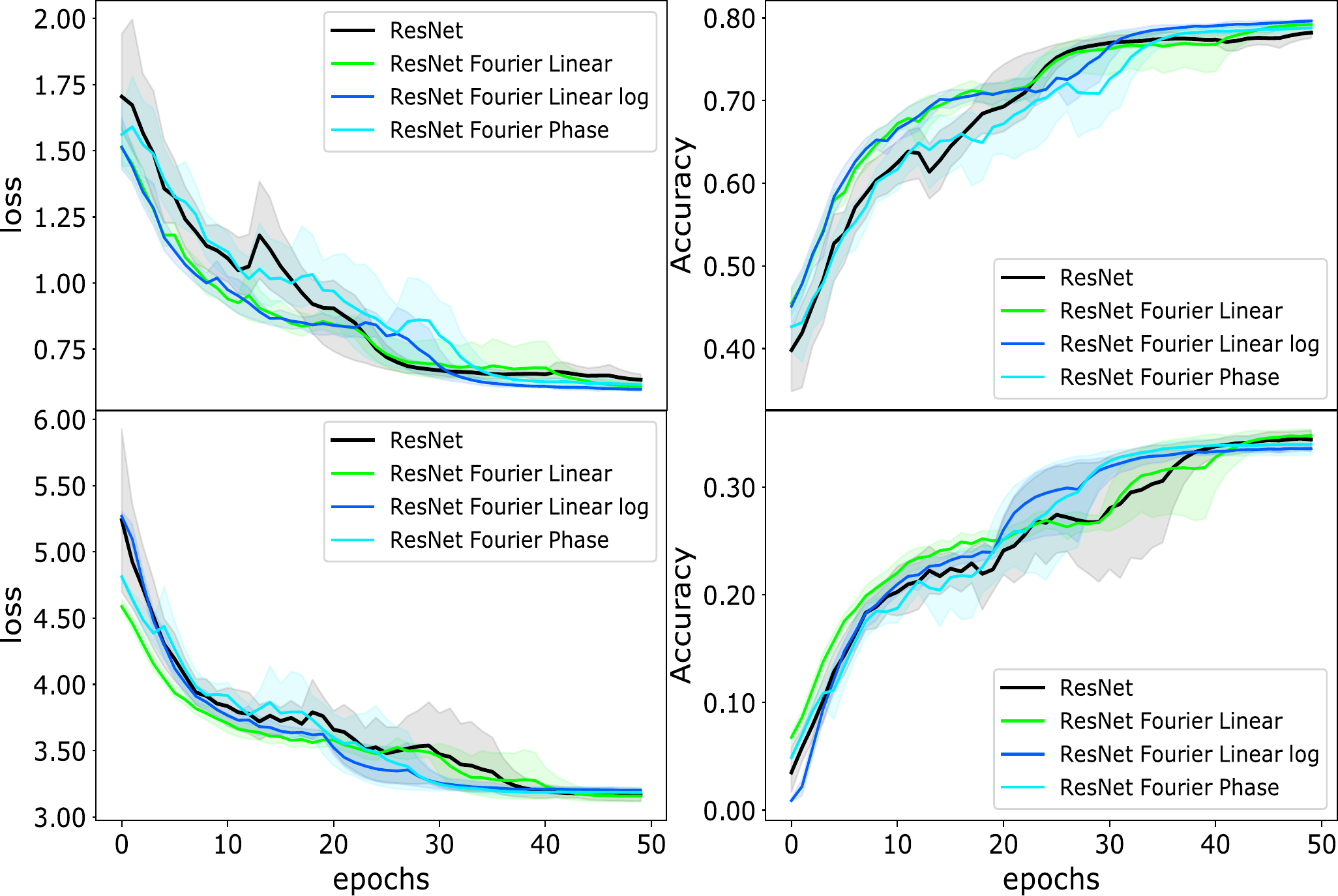}
\caption{\textsc{Control and Large-Scale Experiments}. Loss functions and Accuracy score of different models training processes on CIFAR-10 (top) and Tiny ImageNet (bottom) \emph{validation} sets.}
\label{fig:control_classification}
\end{center}
\end{figure*}

\begin{figure*}[!h]
\begin{center}
\includegraphics[width=\textwidth]{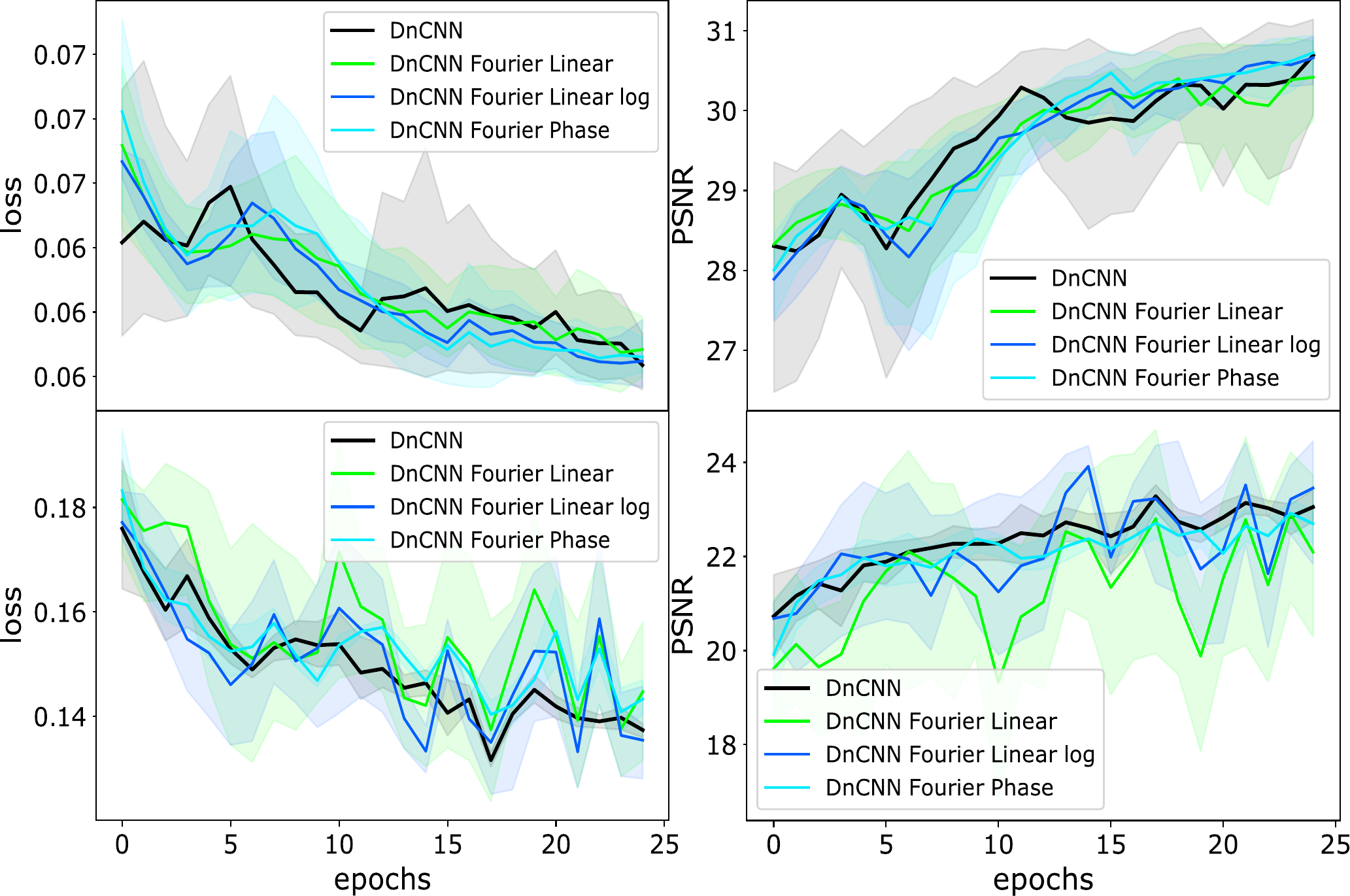}
\caption{\textsc{Control and Large-Scale Experiments}. Loss functions and PSNR metric of different models training processes for Gaussian denoising (top) and erasing corruption (bottom) problems on BUSI \emph{validation} set.}
\label{fig:control_denoising}
\end{center}
\end{figure*}

\paragraph{Phase.} We also looked at the phase adjustment, namely, the rotation of each frequency by the trainable angle. The weight matrix remains exactly the same size as in the \textit{Linear} configuration, but now each weight encodes the angle for which the rotation operator is constructed for each frequency:

\[
\begin{bmatrix} f \\ g \end{bmatrix}_{kl} \longleftarrow \mathbf{R}_{w_{kl}}\begin{bmatrix} f \\ g \end{bmatrix}_{kl} = \begin{bmatrix} \cos{w_{kl}} & -\sin{w_{kl}} \\ \sin{w_{kl}} & \cos{w_{kl}} \end{bmatrix}\begin{bmatrix} f \\ g \end{bmatrix}_{kl}
\]

where $f_{kl} + jg_{kl}$ is $kl$-th frequency in the Fourier transformed image, $w_{kl}$ is $kl$-th weight in trainable weight matrix $W$. The results of such a learning adjustment can be seen in the Figures~\ref{fig:control_segmentation},~\ref{fig:control_classification},~\ref{fig:control_denoising} for segmentation, classification, denoising and erasing experiments, respectively.

\section{Segmentation Models Description}

Segmentation models have been selected among the following popular ones:

\begin{itemize}
    \item U-Net - the first successful architecture in biomedical image segmentation, in that rich feature representation combines lower-level image one using skip connections.
    \item A model with DenseNet encoder, which contains shorter connections between layers close to the input and those close to the output for more accurate and efficient training.
    \item A model with ResNet encoder, where the layers are reformulated as learning residual functions regarding the layer inputs, instead of learning the unreferenced functions.
\end{itemize}

\section{Activation Functions}

The Table~\ref{tab:activations_val} contains the results of the study of the activation function influence in the \textit{General} configuration of the adaptive layer for medical datasets.

\begin{table*}[!ht]
\caption{Validation Dice score for different activation functions on the BUSI and BPUI datasets in the \textit{General} configuration of the adaptive layer. Red color font corresponds to the worst score, and green font corresponds to the best score.}
\label{tab:activations_val}
\vskip -0.2in
\begin{center}
\begin{small}
\begin{tabular}{lcccccc}
\toprule
\multirow{2}{*}{Activation Function} &\multicolumn{3}{c}{{BUSI}} &\multicolumn{3}{c}{{BPUI}} \\
& U-Net & DenseNet & ResNet & U-Net & DenseNet & ResNet \\
\toprule
Sigmoid & \textcolor{red}{{0.70}} & 0.82 & {0.83} & \textcolor{red}{{0.68}} & \textcolor{red}{{0.75}} & \textcolor{red}{{0.72}} \\
\midrule
ReLU & 0.77 & \textcolor{dgreen}{{0.84}} & \textcolor{dgreen}{{0.83}} & \textcolor{dgreen}{{0.75}} & \textcolor{dgreen}{{0.77}} & \textcolor{dgreen}{{0.75}} \\
\midrule
ReLU6 & 0.76 & 0.83 & 0.81 & 0.73 & 0.76 & 0.74 \\
\midrule
Softplus & 0.74 & 0.83 & 0.82 & 0.70 & \textcolor{red}{0.75} & 0.74 \\
\midrule
Tanh & 0.75 & 0.83 & \textcolor{dgreen}{{0.83}} & 0.73 & 0.76 & 0.74 \\
\midrule
Swish ($\beta = 1.0$) & \textcolor{dgreen}{{0.78}} & \textcolor{red}{{0.81}} & \textcolor{red}{{0.80}} & 0.74 & 0.76 & \textcolor{dgreen}{{0.75}} \\
\midrule
Mish & \textcolor{dgreen}{{0.78}} & 0.83 & 0.82 & \textcolor{dgreen}{{0.75}} & \textcolor{dgreen}{{0.77}} & \textcolor{dgreen}{{0.75}} \\
\bottomrule
\end{tabular}
\end{small}
\end{center}
\vskip -0.2in
\end{table*}

\clearpage
\section{Additional Learning Process Results}

\begin{figure*}[!h]
\begin{center}
\includegraphics[width=\textwidth]{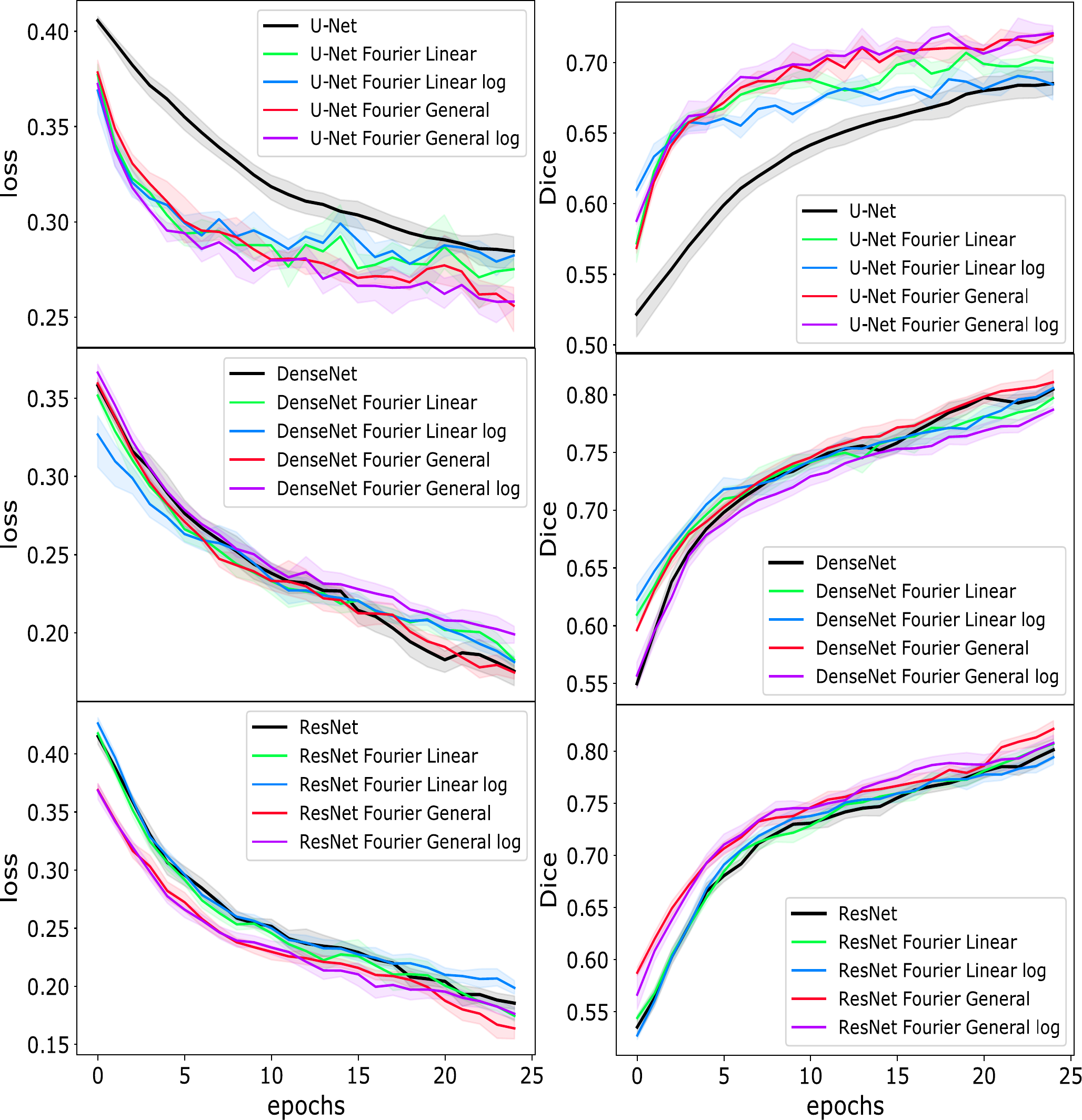}
\caption{\textsc{Segmentation}. Loss functions and Dice coefficients of different models training processes on BUSI \emph{train} set.}
\label{fig:segmentation_BUSI_train}
\end{center}
\end{figure*}

\begin{figure*}[!h]
\begin{center}
\includegraphics[width=\textwidth]{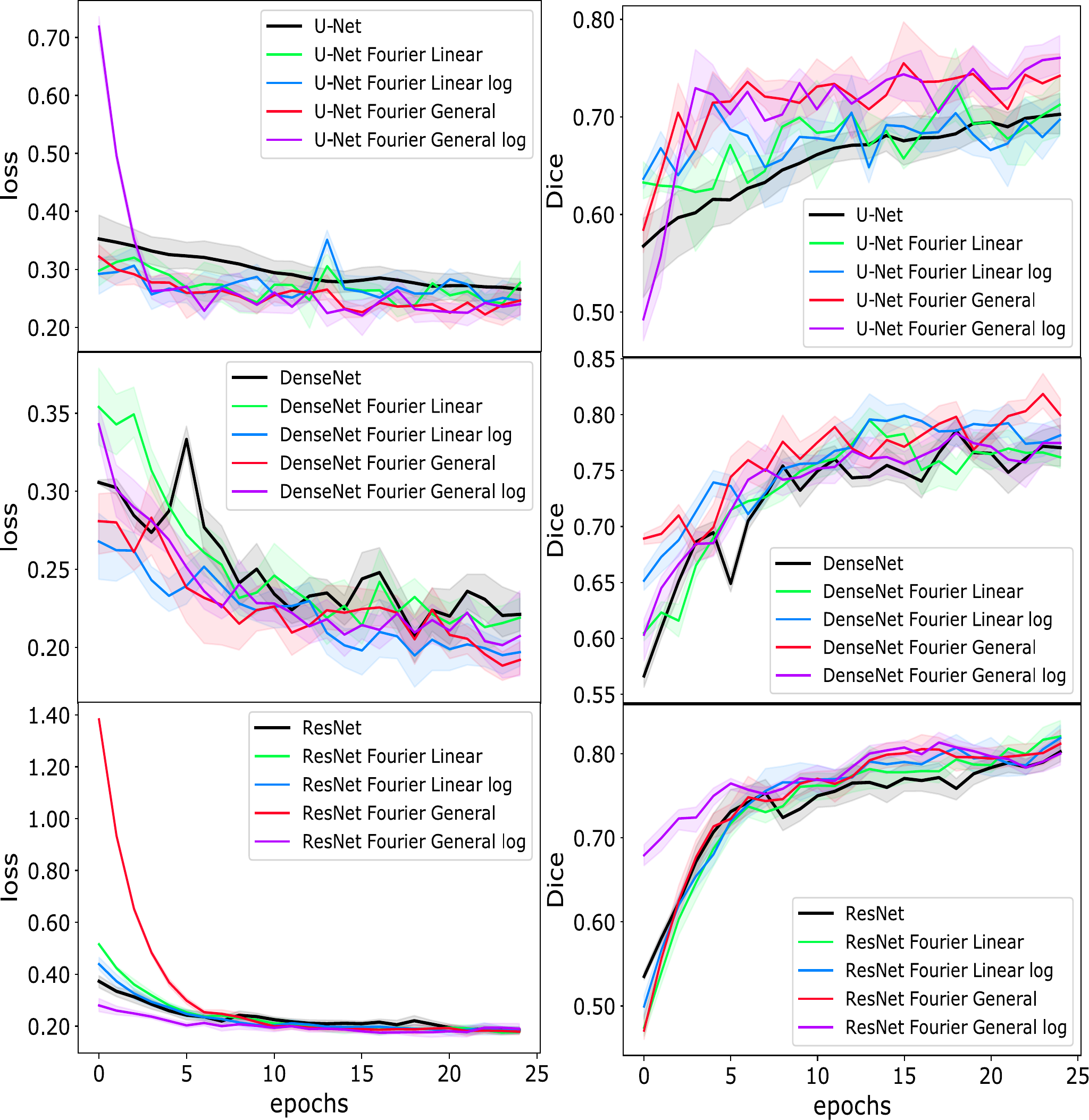}
\caption{\textsc{Segmentation}. Loss functions and Dice coefficients of different models training processes on BUSI \emph{validation} set.}
\label{fig:segmentation_BUSI_val}
\end{center}
\end{figure*}

\begin{figure*}[!h]
\begin{center}
\includegraphics[width=\textwidth]{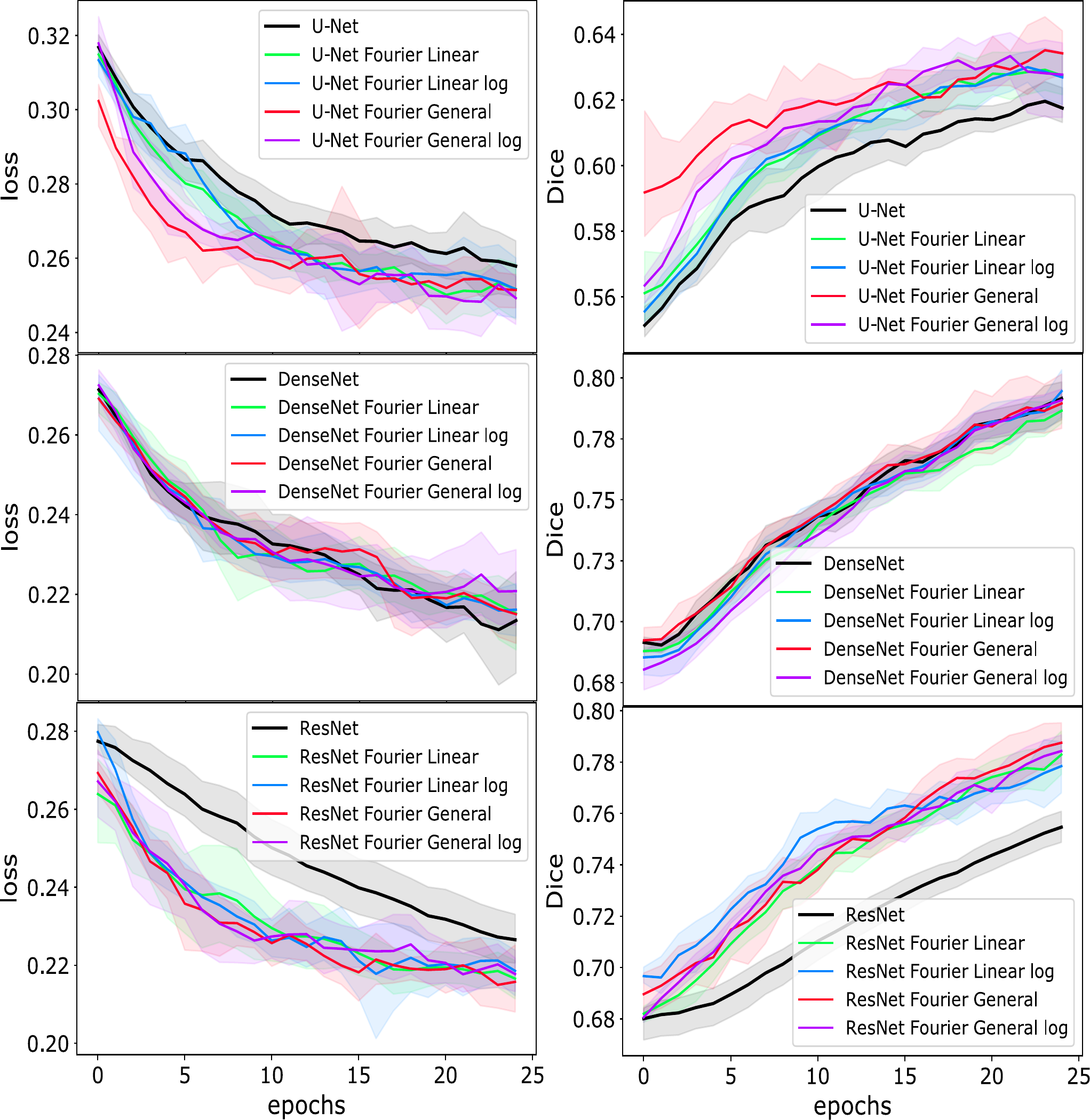}
\caption{\textsc{Segmentation}. Loss functions and Dice coefficients of different models training processes on BPUI \emph{train} set.}
\label{fig:segmentation_BPUI_train}
\end{center}
\end{figure*}

\begin{figure*}[!h]
\begin{center}
\includegraphics[width=\textwidth]{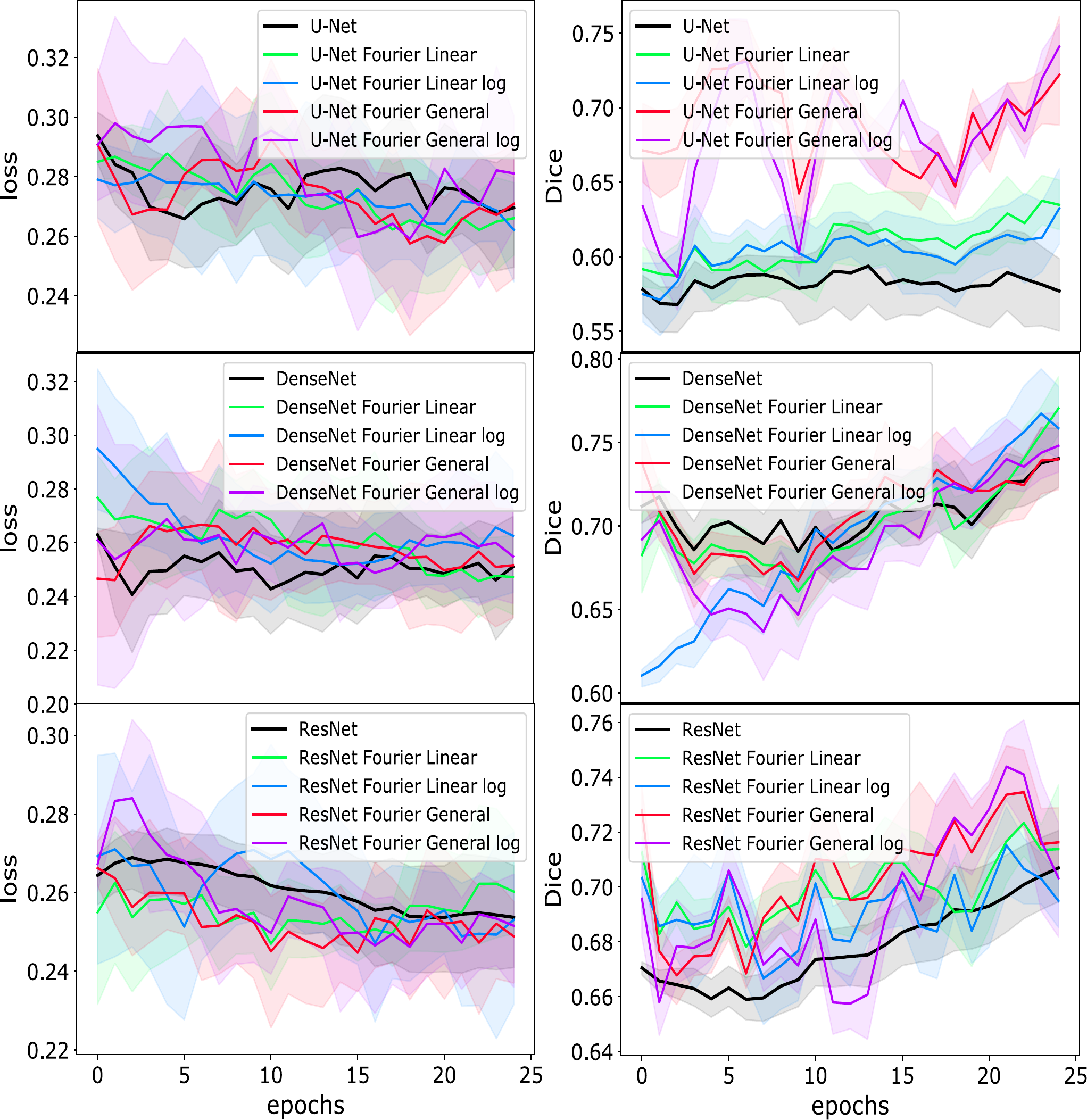}
\caption{\textsc{Segmentation}. Loss functions and Dice coefficients of different models training processes on BPUI \emph{validation} set.}
\label{fig:segmentation_BPUI_val}
\end{center}
\end{figure*}

\begin{figure*}[!h]
\begin{center}
\includegraphics[width=\textwidth]{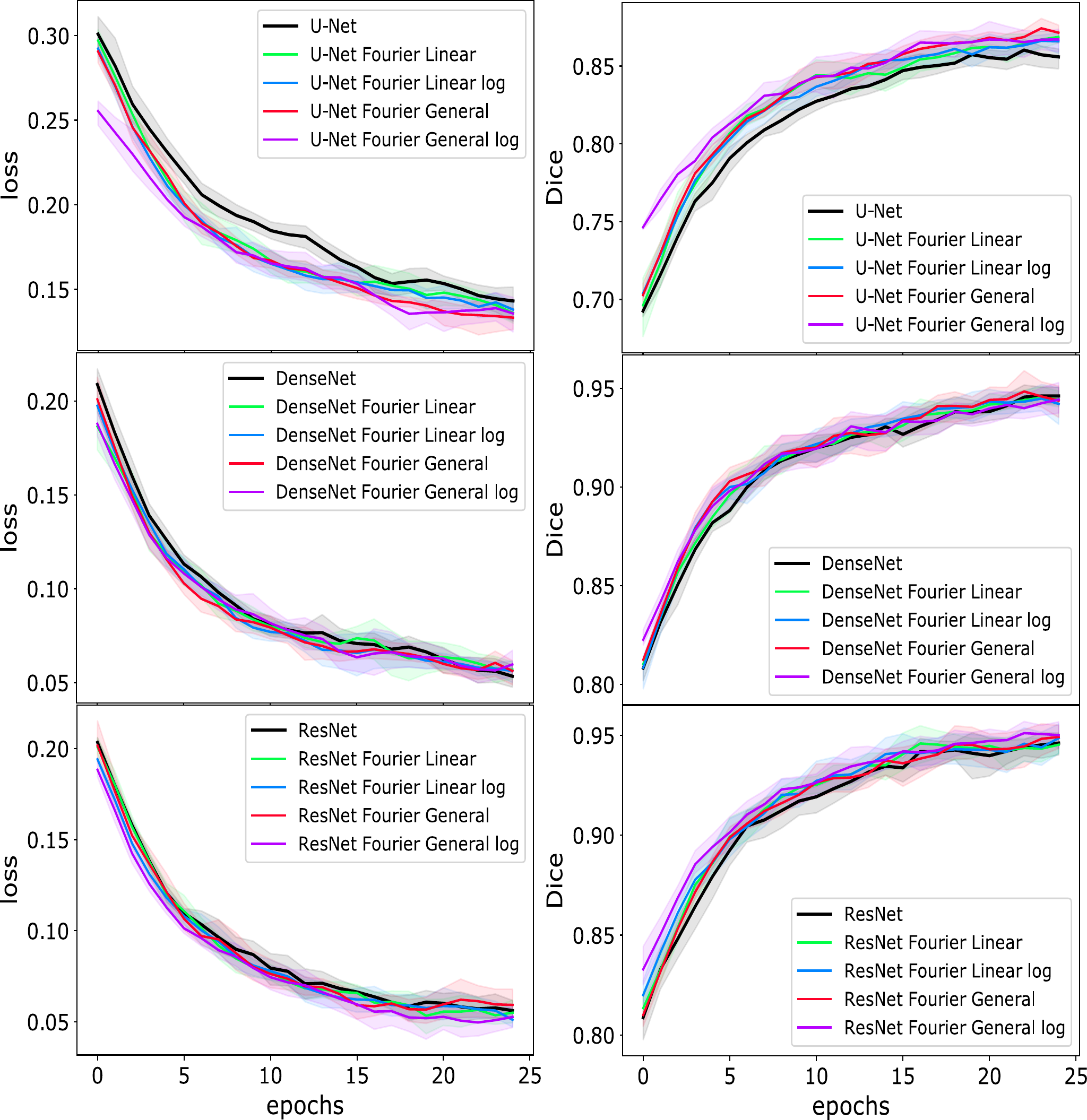}
\caption{\textsc{Segmentation}. Loss functions and Dice coefficients of different models training processes on Caltech Birds (2011) \emph{train} set.}
\label{fig:segmentation_Birds_train}
\end{center}
\end{figure*}

\begin{figure*}[!h]
\begin{center}
\includegraphics[width=\textwidth]{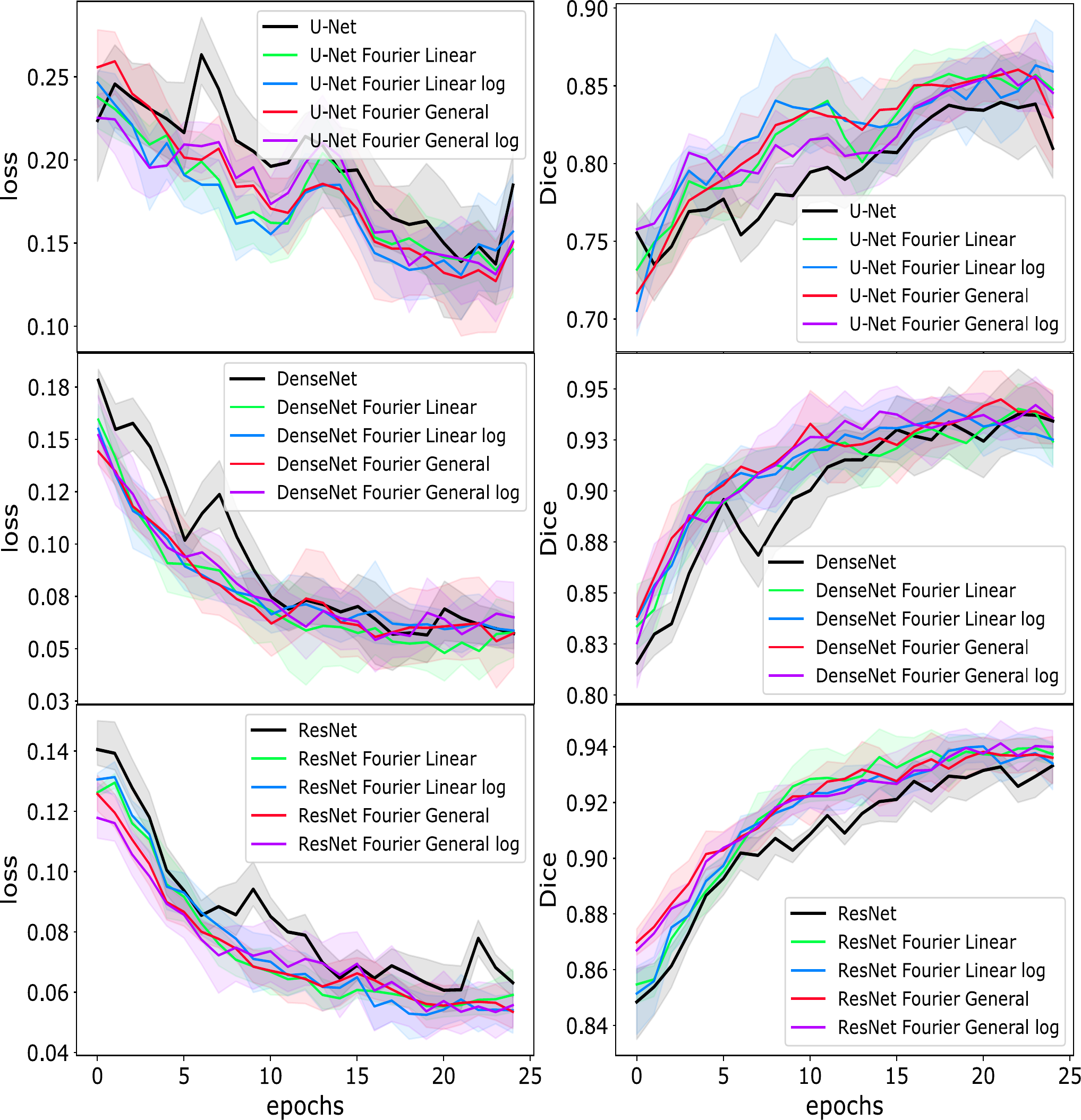}
\caption{\textsc{Segmentation}. Loss functions and Dice coefficients of different models training processes on Caltech Birds (2011) \emph{validation} set.}
\label{fig:segmentation_Birds_val}
\end{center}
\end{figure*}

\begin{figure*}[!h]
\centering
\includegraphics[width=\textwidth]{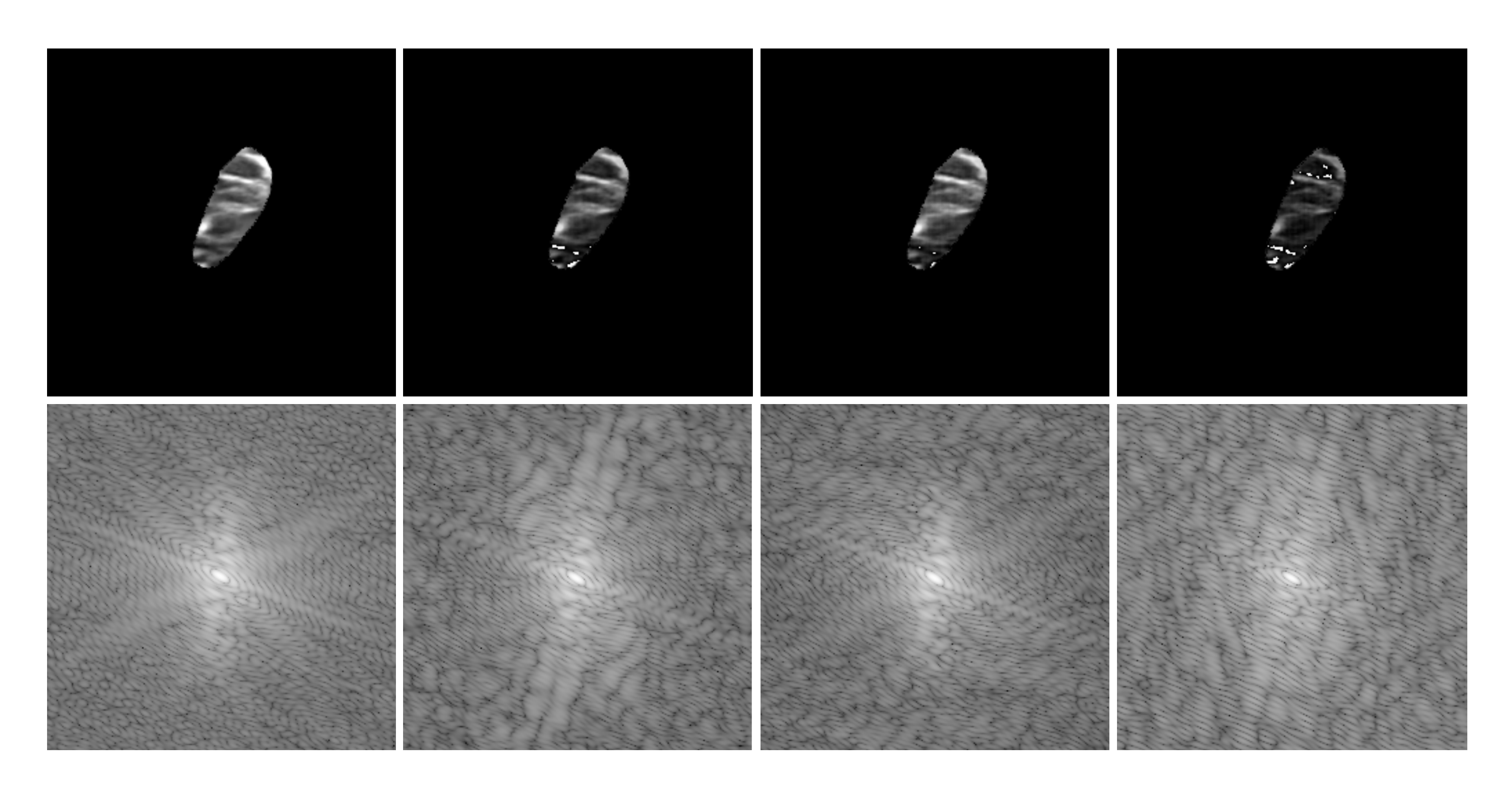}
\caption{\textsc{Segmentation}. Example of trainable filter application to mask area. Initial image, \emph{Linear} filter, \emph{General} filter, \emph{General log} filter.}
\label{fig:filter_to_masks}
\end{figure*}

\begin{figure*}[!h]
\centering
\includegraphics[width=\textwidth]{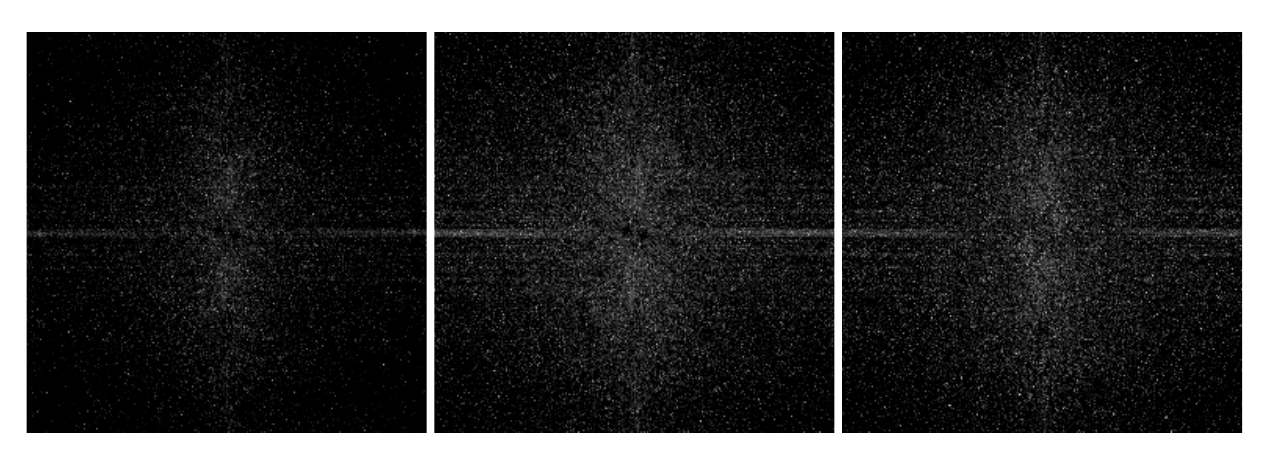}
\caption{\textsc{Segmentation}. Frequency difference of `learnt' filters (left to right: \emph{Linear}, \emph{General}, \emph{General log}) with the initial spectrum.}
\label{fig:amplified_frequencies}
\end{figure*}

\begin{figure*}[!h]
\begin{center}
\includegraphics[width=\textwidth]{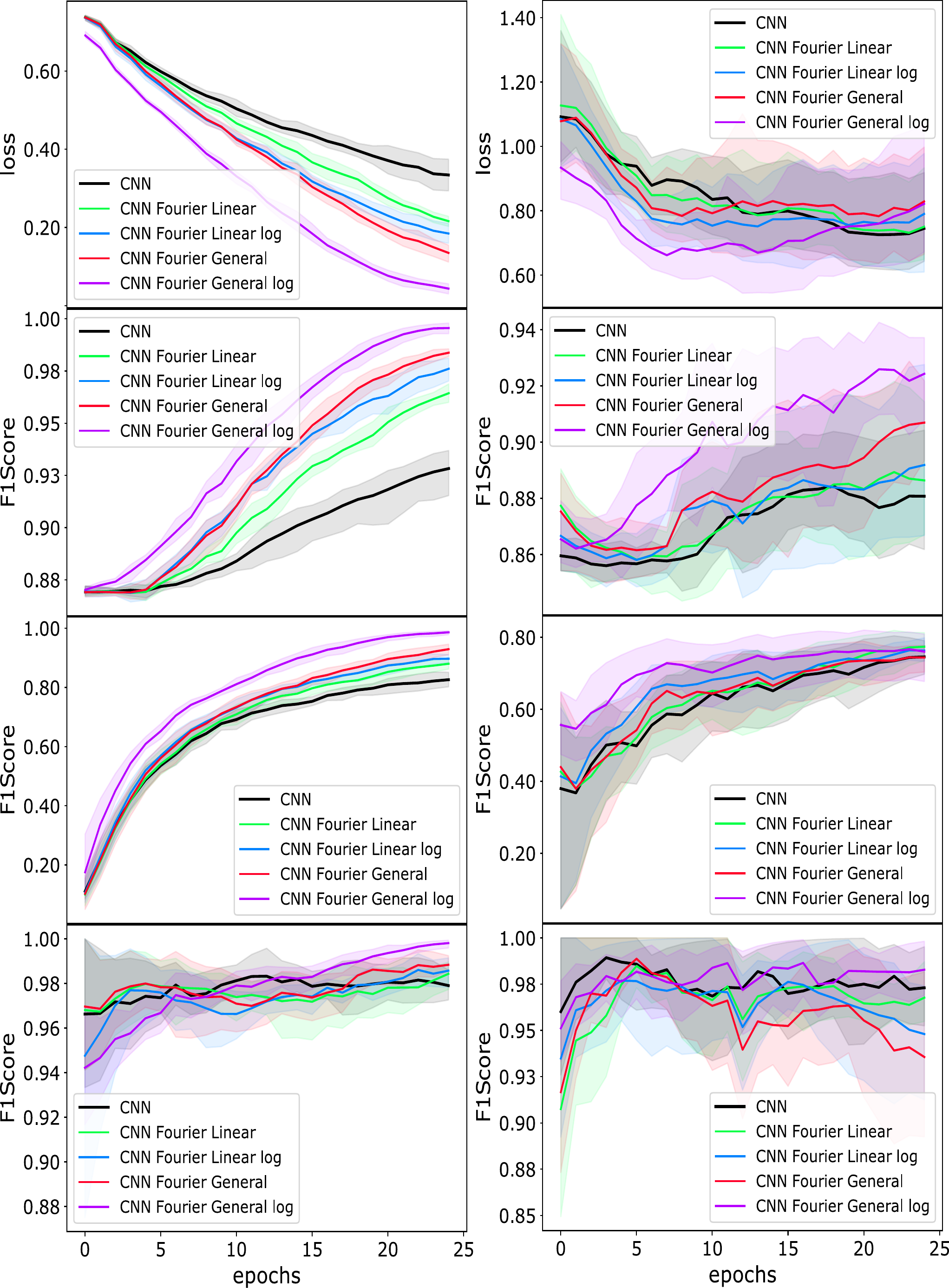}
\caption{\textsc{Classification}. Loss functions and $F_1$-scores of different models training processes on BUSI \emph{train} (left column) and \emph{validation} (right column) sets for normal vs. benign, benign vs. malignant and normal vs. malignant class pairs correspondingly.}
\label{fig:classification_BUSI}
\end{center}
\end{figure*}

\begin{figure*}[!h]
\begin{center}
\includegraphics[width=\textwidth]{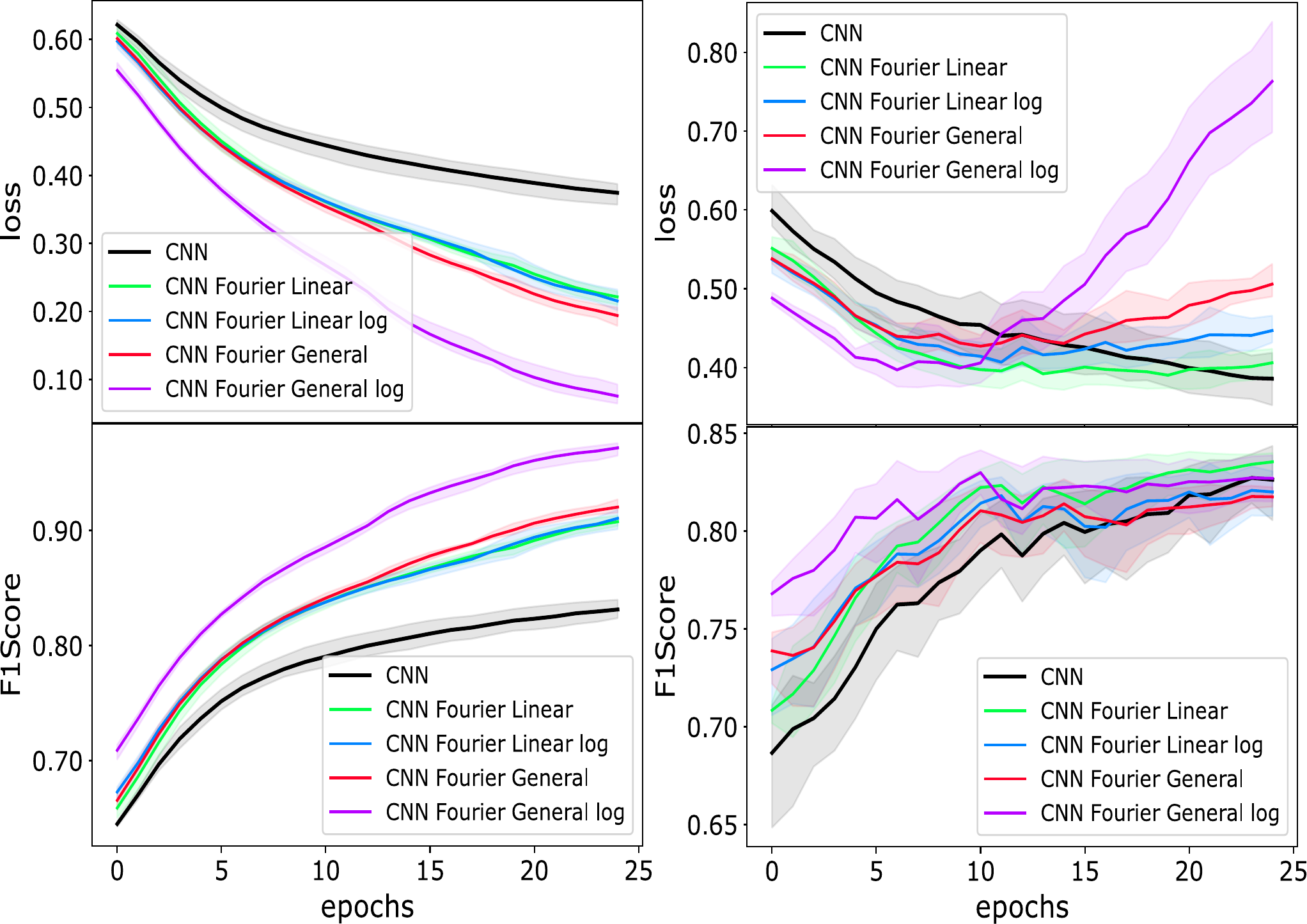}
\caption{\textsc{Classification}. Loss functions and $F_1$-scores of different models training processes on Dogs vs. Cats \emph{train} (left column) and \emph{validation} (right column) sets.}
\label{fig:classification_DvsC}
\end{center}
\end{figure*}

\begin{figure*}[!h]
\begin{center}
\includegraphics[width=\textwidth]{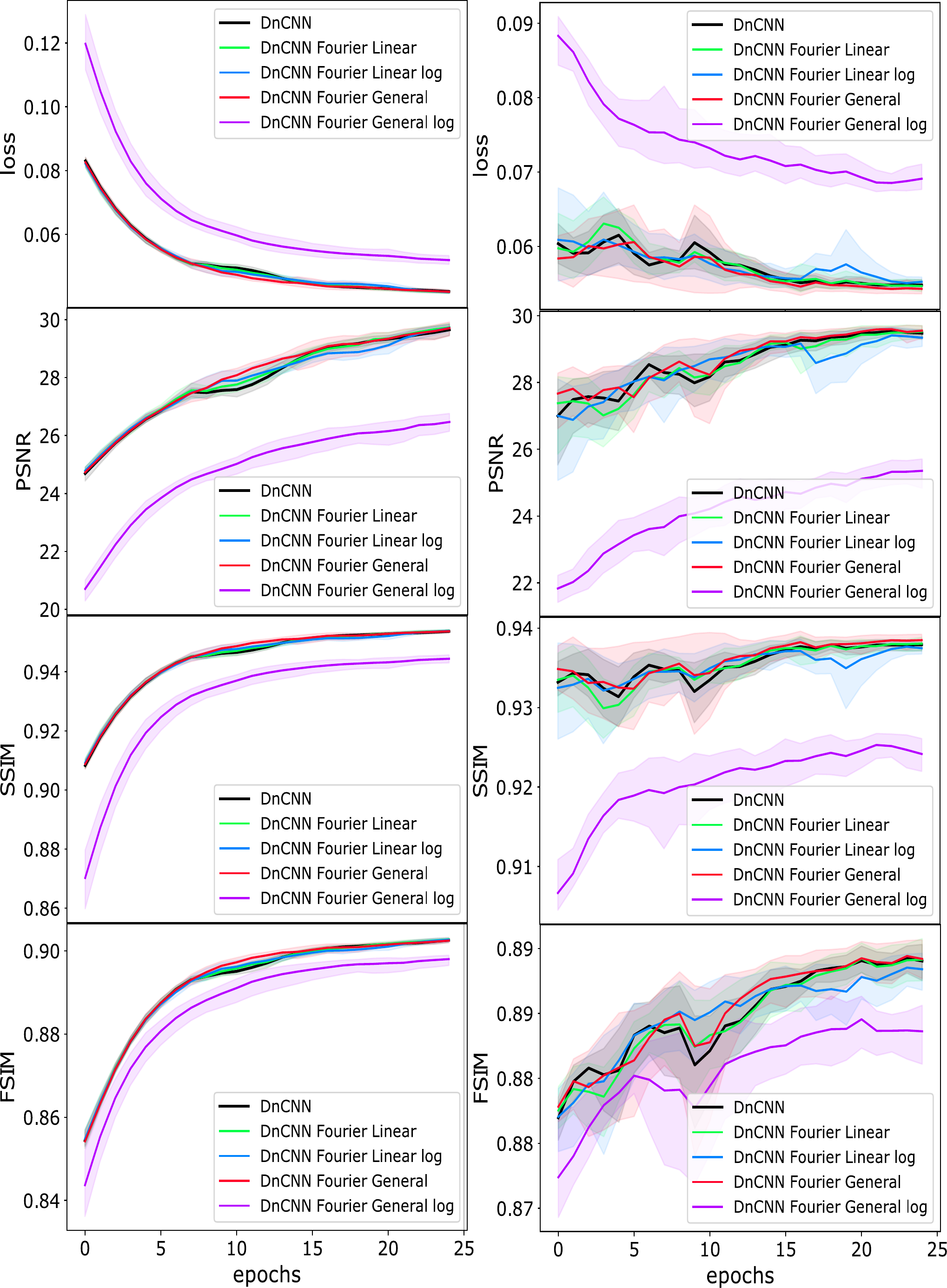}
\caption{\textsc{Denoising}. Loss functions and PSNR, SSIM, FSIM metrics of different models training processes on BUSI \emph{train} (left column) and \emph{validation} (right column) sets.}
\label{fig:denoising_BUSI}
\end{center}
\end{figure*}

\begin{figure*}[!h]
\begin{center}
\includegraphics[width=\textwidth]{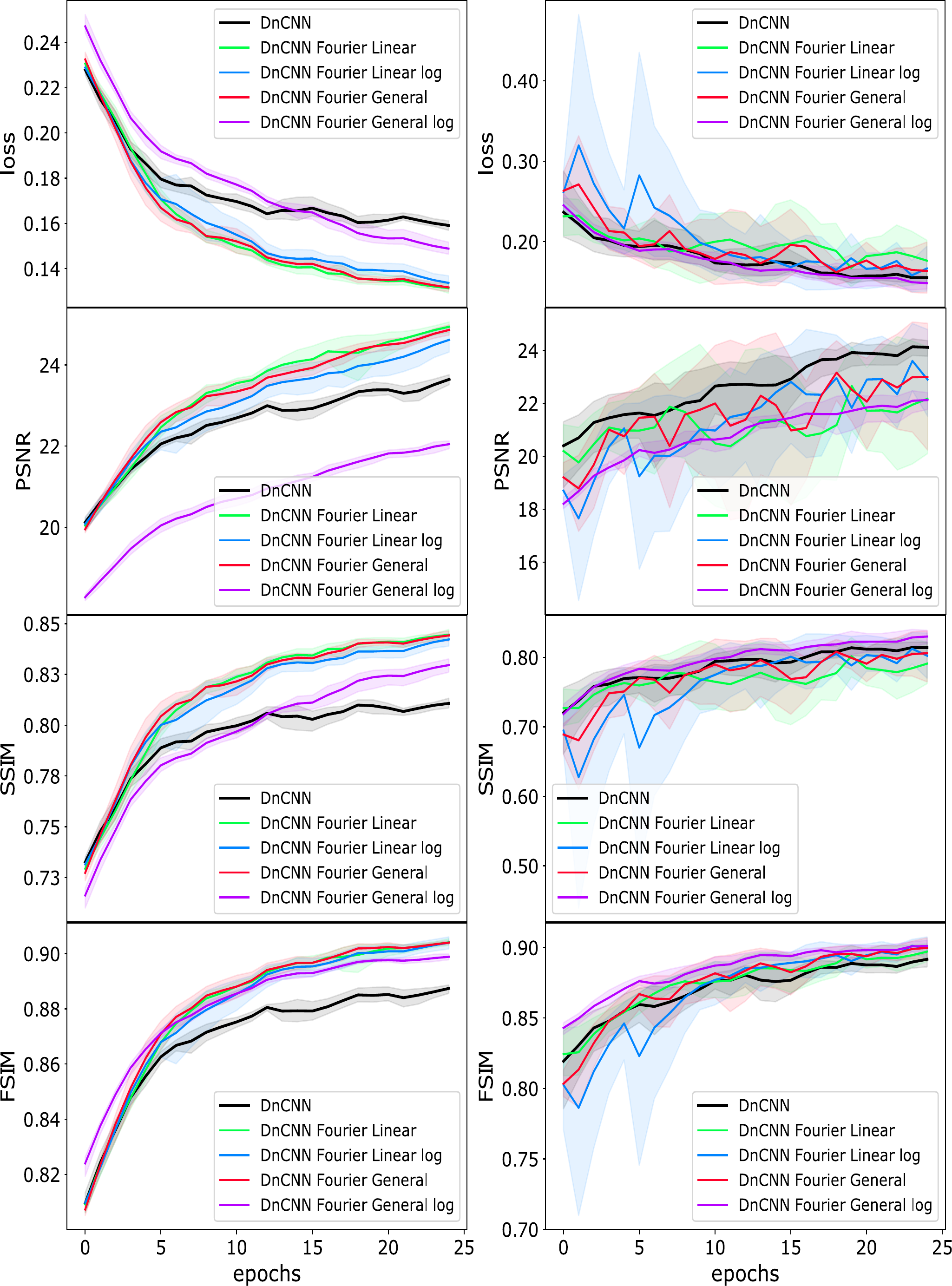}
\caption{\textsc{Erasing}. Loss functions and PSNR, SSIM, FSIM metrics of different models training processes on BUSI \emph{train} (left column) and \emph{validation} (right column) sets.}
\label{fig:erasing_BUSI}
\end{center}
\end{figure*}

\begin{figure*}[!h]
\begin{center}
\includegraphics[width=\textwidth]{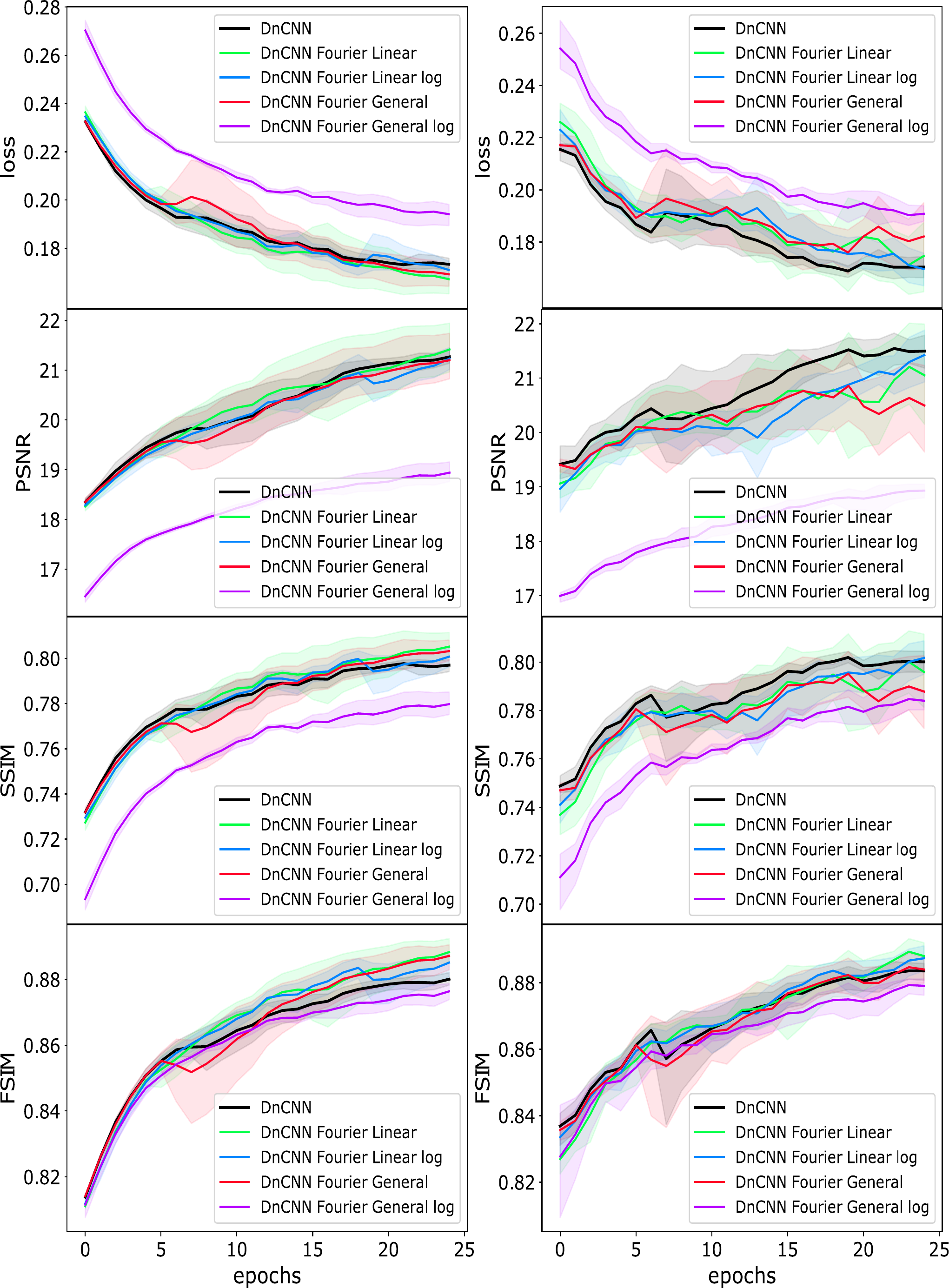}
\caption{\textsc{Erasing}. Loss functions and PSNR, SSIM, FSIM metrics of different models training processes on Dogs vs. Cats \emph{train} (left column) and \emph{validation} (right column) sets.}
\label{fig:erasing_DvsC}
\end{center}
\end{figure*}

As extra information, we provide visualization of learning processes for segmentation (Figures~\ref{fig:segmentation_BUSI_train},~\ref{fig:segmentation_BUSI_val},~\ref{fig:segmentation_BPUI_train},~\ref{fig:segmentation_BPUI_val},~\ref{fig:segmentation_Birds_train},~\ref{fig:segmentation_Birds_val},~\ref{fig:filter_to_masks},~\ref{fig:amplified_frequencies}), classification (Figures~\ref{fig:classification_BUSI},~\ref{fig:classification_DvsC}), denoising (Fig.~\ref{fig:denoising_BUSI}) and erasing (Figures~\ref{fig:erasing_BUSI},~\ref{fig:erasing_DvsC}) tasks.

\end{document}